\documentclass[twocolumn,showpacs]{revtex4}

\usepackage{graphicx}
\usepackage{dcolumn}
\usepackage{amsmath}
\usepackage{amssymb}
\usepackage{color}

\begin{document}

\title{Quantifying Spatiotemporal Chaos in Rayleigh-B\'enard Convection}
\author{A. Karimi}
\affiliation{Department of Engineering Science and Mechanics, Virginia Polytechnic Institute and State University,
 Blacksburg, Virginia 24061}
\author{M. R. Paul}
\affiliation{Department of Mechanical Engineering, Virginia Polytechnic Institute and State University, Blacksburg, Virginia 24061}

\begin{abstract}

Using large-scale parallel numerical simulations we explore spatiotemporal chaos 
in Rayleigh-B\'{e}nard convection in a cylindrical domain with experimentally 
relevant boundary conditions.  We use  the variation of the spectrum of Lyapunov 
exponents and the leading order Lyapunov vector with system parameters to 
quantify states of high-dimensional chaos in fluid convection.  We explore the relationship 
between the time dynamics of the spectrum of Lyapunov exponents and the pattern 
dynamics. For chaotic dynamics we find that all of the Lyapunov exponents 
are positively correlated with the leading order Lyapunov exponent and we 
quantify the details of their response to the dynamics of defects.  The leading order 
Lyapunov vector is used to identify topological features of the fluid patterns that 
contribute significantly to the chaotic dynamics. Our results show a transition 
from boundary dominated dynamics to bulk dominated dynamics as the system size is increased. 
The spectrum of Lyapunov exponents is used to compute the variation of the fractal 
dimension with system parameters to quantify how the underlying high-dimensional 
strange attractor accommodates a range of different chaotic dynamics.
\end{abstract}


\maketitle

\section{Introduction}

At the core of many problems of scientific interest is a spatially extended system 
that is driven far-from-equilibrium to yield spatiotemporal chaos (aperiodic 
dynamics in both space and time)~\cite{cross:1993}. Examples include the dynamics of the 
weather and climate~\cite{lorenz:1963}; fluid turbulence~\cite{holmes:1996}; the 
intricate patterns that occur for reacting, diffusing and advecting chemicals~\cite{nugent:2004}; 
and the transition to chaos in excitable media such as cardiac tissue~\cite{bar:1993}.  It is 
expected for systems such as these that the dimension describing the attractor 
of the dynamics will be very large.  As a result, the powerful ideas of chaotic time series 
analysis~\cite{abarbanel:1996}, as well as geometrical based approaches for estimating 
the dimension~\cite{farmer:1983}, are difficult to apply and are often ineffective.

However, with the advance and availability of sophisticated parallel algorithms 
and supercomputing resources these high-dimensional systems are accessible 
to Lyapunov exponent and Lyapunov vector based diagnostics. Using the standard 
approach~\cite{wolf:1985} of simultaneously evolving the tangent space 
equations with frequent Gram-Schmidt reorthonormalizations allows one to 
compute the spectrum of Lyapunov exponents. With knowledge of the Lyapunov 
exponents the fractal dimension can be estimated using the well known 
Kaplan-Yorke equation~\cite{ott:1993}.

A powerful aspect of this approach is that very large dimensions are now accessible 
with an algorithm that scales readily to parallel computing resources.  Using this 
approach we discuss results for Rayleigh-B\'{e}nard convection which is the 
buoyancy driven fluid convection that occurs in a shallow fluid layer that is 
heated uniformly from below.  Rayleigh-B\'{e}nard convection is a canonical 
system for the study of pattern formation in systems that are driven 
far-from-equilibrium~\cite{cross:1993,bodenschatz:2000}.  The study of 
Rayleigh-B\'enard convection continues to play an important role in 
building our physical understanding of the complex dynamics that occur 
in driven spatially-extended systems.  

The desire for a quantitative understanding of high-dimensional spatiotemporal 
chaos for experimentally accessible systems is an important challenge.  
In this paper we discuss results for experimentally accessible conditions 
with fractal dimensions as large as 50.  To the best of our knowledge this 
represents the highest dimension dynamics that have been explored 
using Lyapunov based diagnostics for laboratory conditions. Knowledge 
of the fractal dimension can be used to provide fundamental insights into the 
underlying chaotic dynamics. The numerical value of the fractal dimension 
provides an estimate for the number of chaotic degrees of freedom that are 
active in the system~\cite{farmer:1983}.  Given the number of chaotic 
degrees of freedom that describe the dynamics one can construct estimates 
for the length scales of these degrees of freedom on average. In addition, the 
variation of the fractal dimension with changing system parameters allows one to 
probe quantitatively how the attractor accommodates different dynamics.

In the literature there are a number of new insights provided by the 
study of fluid convection using information gained from computing 
Lyapunov based diagnostics.  Egolf \textit{et al.}~\cite{egolf:2000} 
demonstrated that Rayleigh-B\'enard convection exhibited extensive 
chaos for large periodic domains with aspect ratios $48 \le \Gamma \le 64$ 
where $\Gamma = L/d$, $L$ is the side length of the domain,  
and $d$ is the depth of the fluid layer.  In this study the system parameters 
were chosen to yield the spiral defect chaos state~\cite{morris:1993}.  
The spatiotemporal dynamics of the leading order Lyapunov vector was 
studied and was found to be largest in regions where roll pinch-off events 
were occurring.  It was suggested that the dynamics of these local defects 
were contributing significantly to the disorder as opposed to the visually 
striking spiral structures.

Scheel and Cross~\cite{scheel:2006} used the leading-order Lyapunov 
exponent and Lyapunov vector to perform a careful study of the 
time-periodic and chaotic dynamics that occur in a small cylindrical 
convection layer with $\Gamma =  5$ (where $\Gamma = r_0/d$ 
and $r_0$ is the radius of the convection domain). They conclude 
that repeating local defect dynamics involving roll pinch-off events 
contribute significantly to the short-time Lyapunov exponent without 
affecting the long-time Lyapunov exponent. Interestingly, they find that 
the non-repeating roll pinch-off events are what contribute significantly 
to the long-time Lyapunov exponent.  This raises several interesting 
questions. How does the leading order Lyapunov exponent discern 
between repeating and non-repeating events? How do the other Lyapunov 
exponents in the Lyapunov spectrum respond to these events? In this 
paper we will shed some further insight upon these questions.

Paul \textit{et. al}~\cite{paul:2007} computed the spectrum of Lyapunov 
exponents for chaotic convection in cylindrical domains for aspect 
ratios $4.72 \lesssim \Gamma \le 15$. It was determined that Rayleigh-B\'enard 
convection was extensively chaotic for $\Gamma \gtrsim 7$. Jayaraman 
\textit{et al.}~\cite{jayaraman:2006} explored the leading-order Lyapunov exponent 
and Lyapunov vector for the domain chaos state that occurs for Rayleigh-B\'enard 
convection in a rotating domain.  An interesting feature of domain chaos is 
the presence of propagating fronts as well as localized defect structures. A 
careful study revealed that not all defect structures contributed equally to 
the leading order Lyapunov exponent, a result that is in agreement with the 
findings of Scheel and Cross~\cite{scheel:2006} for the spiral defect chaos state.

In this paper we present a detailed study of chaotic Rayleigh-B\'enard convection 
using diagnostics based on the spectrum of Lyapunov exponents and Lyapunov 
vectors for a range of experimentally relevant conditions. In Section II we describe 
the numerical approach used to compute the flow fields, Lyapunov exponents, and 
Lyapunov vectors.  In Section III we discuss the dynamics of the Lyapunov exponents, 
the spatiotemporal features of the leading order Lyapunov vector, and the variation of 
the fractal dimension with system parameters. Lastly, in Section IV we present our 
concluding remarks. 

\section{Approach}

\subsection{Rayleigh-B\'{e}nard Convection}

Rayleigh-B\'{e}nard convection is the buoyancy-driven motion that results when a 
thin layer of fluid is heated uniformly from below.  The fluid motion is described by 
the Boussinesq equations, 
\begin{eqnarray}
	\sigma^{-1} \left(\partial_t + \mathbf{u}\cdot\mathbf{\nabla} \right) \mathbf{u} &=& -\mathbf{\nabla} p + \nabla^{2}\mathbf{u}+ R T \hat{\mathbf{z}},  \label{eq:boussinesq1}\\
	\left(\partial_t + \mathbf{u} \cdot \mathbf{\nabla} \right) T &=& \nabla^{2} T,  \label{eq:boussinesq2}\\
	\mathbf{\nabla} \cdot \mathbf{u} &=& 0,
 \label{eq:boussinesq3}
\end{eqnarray}
where $\hat{\mathbf{z}}$ is a unit vector in the $z$-direction that opposes gravity, $\sigma$ 
is the Prandtl number, $R$ is the Rayleigh number, $ \mathbf{u}$ is the fluid velocity, $p$ is 
the pressure, and $T$ is the temperature.  The equations are nondimensionalized using 
the layer depth $d$ for the length scale, the vertical diffusion time for heat $d^2/\alpha$ 
where $\alpha$ is the thermal diffusivity for the time scale, and the constant temperature 
difference between the bottom and top plates $\Delta T$ as the temperature scale.

The no-slip boundary condition is applied to all material surfaces
\begin{equation}
\mathbf{u} = 0
\end{equation}
and the lateral side-walls of the cylindrical domain are assumed to be perfectly conducting
\begin{equation}
T(z) = 1 - z.
\end{equation}
The Rayleigh number,
\begin{equation}
R = \frac{g \beta \Delta T d^3}{\nu \alpha}
\end{equation}
is the control parameter that is most often varied in experiment.  Small values of $R$ 
correspond to simple, often time-independent flows; intermediate values of $R$ correspond 
to complex chaotic flows as studied here; and large values of $R$ correspond to 
strongly driven turbulent flows~\cite{ahlers:2009}.  It will be convenient to use 
the reduced Rayleigh number $\epsilon = (R-R_c)/R_c$ where $R_c=1707.76$ is the 
critical Rayleigh number for an infinite layer of fluid. The Prandtl number,
\begin{equation}
\sigma = \frac{\nu}{\alpha}
\end{equation}
is the ratio of momentum and thermal diffusivities.  The magnitude of the 
Prandtl number is inversely related  to the strength of the mean flow~\cite{chiam:2003}. 
The mean flow is a weak but long-range flow field that originates from the Reynolds 
stress term and is driven by roll curvature, roll compression, and gradients in the convection 
amplitude~\cite{greenside:1988}.  The mean flow is very difficult to measure 
experimentally~\cite{croquette:1986,pocheau:1987} and has a dramatic effect upon 
the linear stability of the convection rolls~\cite{newell:1990:prl,newell:1990:jfm}. 
Its importance is not due to its strength, but because it is a nonlocal effect acting over large distances 
(many roll widths) and advects the pattern~\cite{cross:1984}.

The aspect ratio of the domain $\Gamma$ is a measure of the spatial extent of the system. 
The dynamics of the flow field depends strongly upon the aspect ratio of the fluid 
layer~\cite{ahlers:1974}.  For small domains the sidewalls tend to frustrate the dynamics 
due to the tendency of the convection rolls to approach a sidewall with the roll axis 
perpendicular to the boundary.  In cylindrical domains this leads to the presence of wall 
foci which can penetrate several roll wavelengths into the domain. As the aspect ratio 
increases the influence of the sidewalls diminishes.

\subsection{Computing the Lyapunov Exponents and Lyapunov Vectors}

We compute the spectrum of Lyapunov exponents $\lambda_k$ using the standard 
procedure described in detail in Ref.~\cite{wolf:1985}. For each exponent a set of equations 
linearized about Eqs.~(\ref{eq:boussinesq1})-(\ref{eq:boussinesq3}) are evolved 
simultaneously to yield the dynamics of perturbations arbitrarily close to the full 
nonlinear system. These tangent space equations are:
\begin{eqnarray}
&& \sigma^{-1} \left( \partial_t \delta \mathbf{u}^{(k)} + \mathbf{u} \cdot \mathbf{\nabla} \delta \mathbf{u}^{(k)} + \delta \mathbf{u}^{(k)} \cdot \mathbf{\nabla} \mathbf{u} \right)  = -\mathbf{\nabla} \delta p^{(k)} \nonumber\\
&& \qquad \qquad \qquad \qquad \qquad \quad+ \nabla^{2}\delta \mathbf{u}^{(k)} + R \delta T^{(k)} \hat{\mathbf{z}},  \label{eq:tangent1} \\
&& \partial_t \delta T^{(k)} + \mathbf{u} \cdot \mathbf{\nabla} \delta T^{(k)} + \delta \mathbf{u}^{(k)} \cdot \mathbf{\nabla} T= \nabla^2 \delta T^{(k)}, \label{eq:tangent2} \\
&& \mathbf{\nabla} \cdot \delta \mathbf{u}^{(k)} = 0. 
\label{eq:tangent3}
\end{eqnarray}
which can be written as,
\begin{equation}
	\frac{d \delta \mathbf{H}^{(k)}}{dt} = \mathbf{J}\left[ \mathbf{H}(t) \right] \delta \mathbf{H}^{(k)},
\label{eq:pertshort}
\end{equation}
where $\mathbf{H}(t)=[\mathbf{u},T]$ and $\delta \mathbf{H}^{(k)}(t)=\left[ \delta \mathbf{u}^{(k)}(t),\delta T^{(k)}(t) \right]$. 
For incompressible fluid flow the pressure is implicitly determined by the requirement of the conservation 
of mass. As a result, the vectors $\mathbf{H}(t)$ and $\delta \mathbf{H}^{(k)}(t)$ do not include $p$ and 
$\delta p$, respectively. In our notation, $\mathbf{J} = d \mathbf{F} /d \mathbf{H}$ where $\mathbf{J}$ is the Jacobian 
of the flow that results when rewriting Eqs.~(\ref{eq:boussinesq1})-(\ref{eq:boussinesq3}) as 
$d \mathbf{H}(t)/dt = \mathbf{F(H)}$.  The boundary conditions for the perturbation equations 
are $\delta \mathbf{u}^{(k)}=\mathbf{0}$ and $\delta \mathbf{T}^{(k)}=0$ at all material 
walls. 

The perturbations are reorthonormalized using a Gram-Schmidt procedure after a time 
$t_N$ to yield the magnitude of their growth $\|\delta \mathbf{H}^{(k)}(t_N)\|$ where the 
normalization is defined over the interior volume $V$ as,
\begin{equation}
\|\delta \mathbf{H}^{(k)}(t)\| = \sqrt{\frac{1}{V} \int_V \left[ \delta \mathbf{u}^{(k)}(t)^2 + \delta T^{(k)}(t)^2 \right] dV}.
\label{eq:normal}
\end{equation}
Each reorthonormalization yields a value of the instantaneous Lyapunov exponent,
\begin{equation}
\tilde{\lambda}_k=\frac{1}{t_N} \ln \|\delta \mathbf{H}^{(k)}(t_N)\|.
\label{eq:inst}
\end{equation}
This is repeated and the average value of $\tilde{\lambda}_k$ yields the finite time Lyapunov exponent 
\begin{equation}
\lambda_k=\frac{1}{N_t} \sum_{i=1}^{N_t} \tilde{\lambda}_k
\label{eq:lambda}
\end{equation}
where $N_t$ is the number of reorthonormalizations performed. The limit $N_t \rightarrow \infty$ 
yields the infinite-time Lyapunov exponent. 

The leading-order exponent $\lambda_1$ describes the growth of the line separating two trajectories in phase 
space, $\lambda_1+\lambda_2$ describes the growth of a two-dimensional area of initial conditions, and 
$\sum_{i=1}^{N} \lambda_i$ describes the growth of an $N$-dimensional ball of initial conditions. The 
exact number of exponents required for the sum to vanish corresponds to the dimension of the ball of 
initial conditions that will neither grow nor shrink under the dynamics and is referred to as 
Lyapunov or fractal dimension $D_\lambda$. Given only the Lyapunov exponents, $D_\lambda$ can 
be determined from the Kaplan-Yorke formula,
\begin{equation}
 D_\lambda=K+\frac{S_K}{|\lambda_{K+1}|},
 \label{eq:D}
\end{equation}
where $K$ is the largest $n$ for which $S_n=\sum_{i=1}^n \lambda_i>0$~\cite{cross:1993,ott:1993}. The 
value of $D_\lambda$ is the minimum number of active degrees of freedom that contribute to the chaotic 
dynamics~\cite{farmer:1983}. 

To solve the system of equations given by Eqs.~(\ref{eq:boussinesq1})-(\ref{eq:boussinesq3}) 
and Eqs.~(\ref{eq:tangent1})-(\ref{eq:tangent3}) we used a highly efficient, parallel spectral 
element code developed to solve the Boussinesq equations. This code has been 
used in a number of numerical explorations of Rayleigh-B\'enard convection that 
have been discussed in the literature~(c.f.~\cite{paul:2001,paul:2002:physd,scheel:2006,jayaraman:2006,chiam:2003,paul:2007,duggleby:2010}).  The underlying numerical approach is discussed 
in Refs.~\cite{fischer:1994,fischer:1997} and a discussion of its application to 
Rayleigh-B\'enard convection can be found in Ref.~\cite{paul:2002:physd}.  

In our numerical simulations, we begin from a small random perturbation on the order of $10^{-3}$ to the linear 
conduction temperature profile with zero velocity field.  The initial conditions for the tangent 
space equations are zero perturbation velocity and a random temperature perturbation with 
a magnitude on the order of $10^{-5}$. A typical value of the numerical time step is 
$\Delta t = 10^{-3}$ and we perform a Gram-Schmidt reorthonormalization every 10 time 
steps.  Within each spectral element we have used 11$^{\text{th}}$ order polynomials to 
represent the field variables.

 Over the course of this work we have performed numerous tests by varying the numerical 
parameters used in the code to ensure the validity of our numerical results.  In particular, we have 
performed simulations for varying time steps and spatial discretizations to ensure that our 
results for the Lyapunov-based diagnostics are accurate and reproducible. For a typical 
numerical simulation we integrate the equations for approximately 15 horizontal diffusion times 
to allow for initial transients to decay. We then use the numerical data from the latter half 
of the simulation to compute the Lyapunov diagnostics 
that we report here. Where possible we have included error bars in our results to reflect the 
variation in the quantities presented based upon our numerical results.  

\begin{figure}[tbh]
\begin{tabular}{cc}
\includegraphics[width=2.5in]{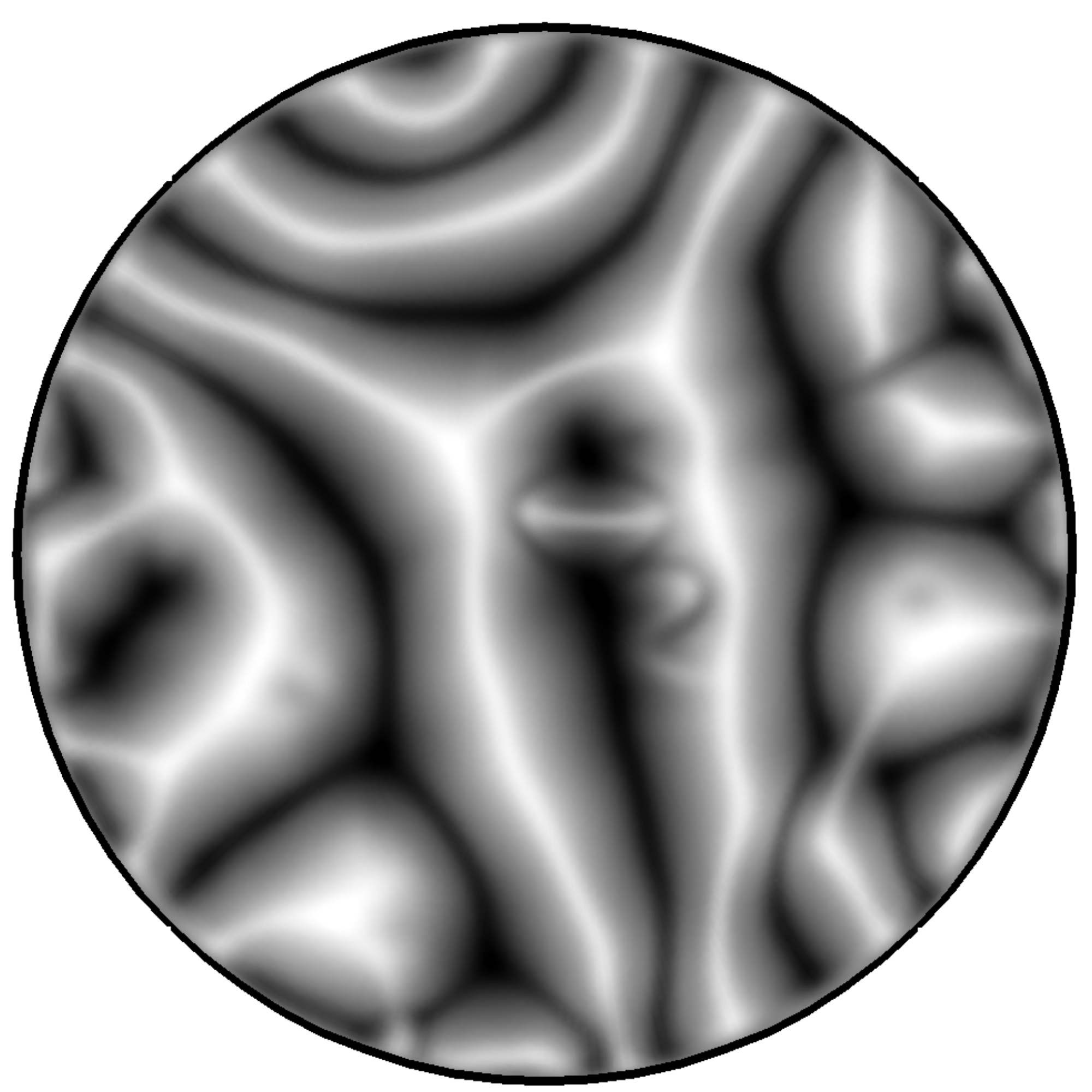}
\end{tabular}
\caption{A spatiotemporally chaotic flow field for $\epsilon=4.27$, $\sigma=1$, 
and $\Gamma=10$. Contours are shown of the temperature field at a 
mid-plane slice where $z=1/2$. Light regions are hot rising fluid and dark regions 
are cool falling fluid. This flow field image is at time $t=610.5$.} 
\label{fig:r9000}
\end{figure}
\begin{figure}[tbh]
\begin{tabular}{cc}
\includegraphics[width=3.5in]{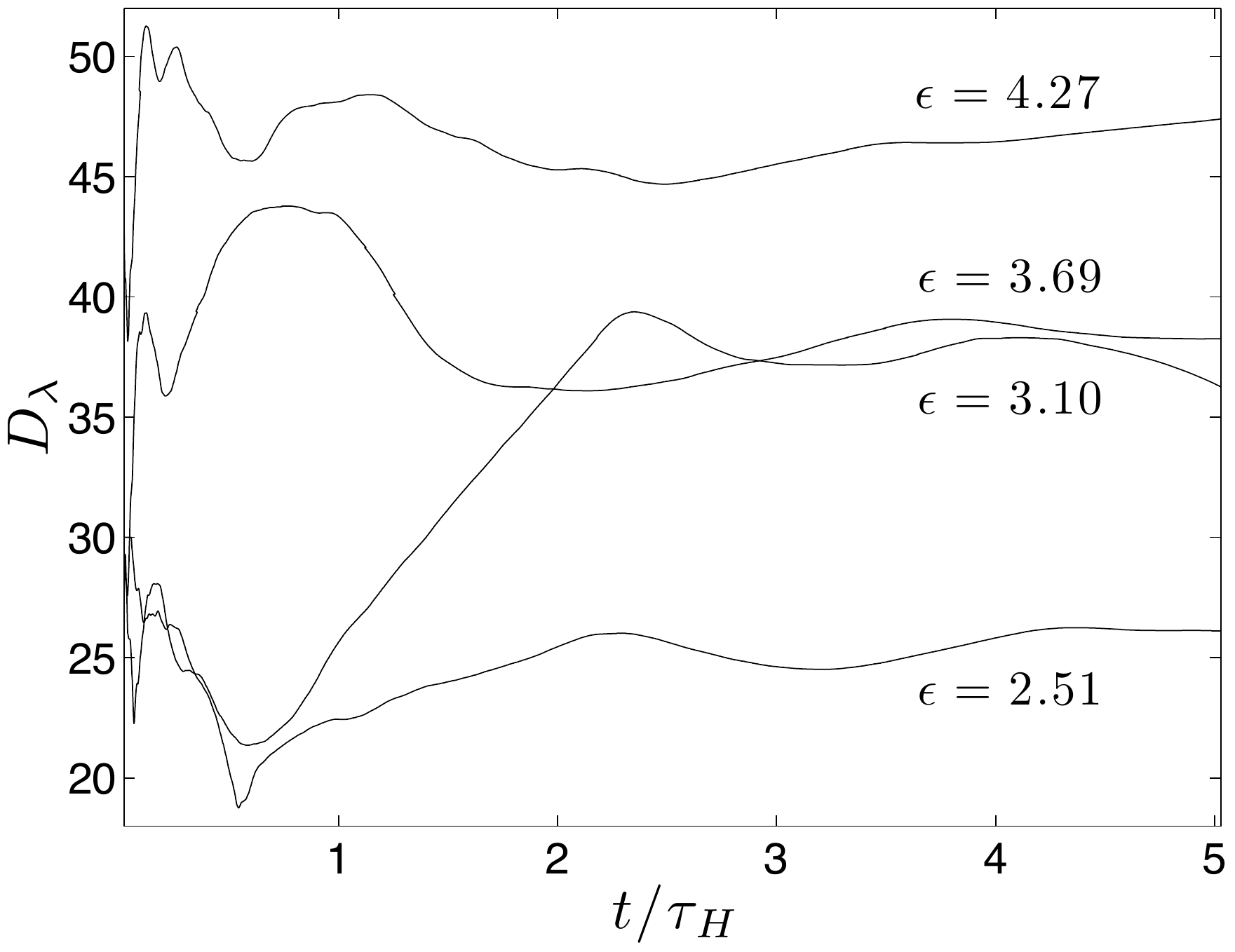} \\
\end{tabular}
\caption{The convergence of the fractal dimension $D_\lambda$ in time. Results are 
shown for 4 different values of the reduced Rayleigh number $\epsilon$ where 
$\Gamma=10$ and $\sigma=1$. The time scale has been normalized by the 
horizontal diffusion time for heat, $\tau_H=\Gamma^2$. The convergence is quite slow and 
remains noisy over the entire range shown.} 
\label{fig:convergence}
\end{figure}

\section{Discussion}

A typical chaotic flow field pattern from our numerical simulations is shown in Fig.~\ref{fig:r9000}. 
The contours of the temperature field are shown at mid-depth where light regions 
are hot rising fluid and dark regions are cool falling fluid.  The fractal dimension $D_\lambda$ of this flow 
field is approximately 50. The convergence of $D_\lambda$ in time is shown in 
Fig.~\ref{fig:convergence} for a range of reduced Rayleigh numbers.  To emphasize the 
slow and noisy convergence the time axis has been normalized by the nondimensional 
horizontal diffusion time $\tau_H = \Gamma^2$ which represents the time required 
for heat to diffuse from the center of the domain to the sidewall.  Time scales on the 
order of the horizontal diffusion time for heat and longer have been shown to describe 
the duration required for large aspect ratio convecting systems to reach dynamics 
that are independent of initial transients~\cite{cross:1984}.

\subsection{The Dynamics of the Lyapunov Exponents}

We are interested in understanding how the time dynamics of the Lyapunov exponents 
relate to the dynamics of the flow field.  Only the leading order Lyapunov vector is pointing 
in a physically relevant direction due to the Gram-Schmidt reorthonormalizations that 
are used in their computation.  The magnitude of the Lyapunov exponents are 
not affected by this and the variation of their magnitude in time provides insight into 
the underlying dynamics.  For example, it would be useful to know if the different 
exponents exhibit different dynamics that could be related to features of the pattern 
dynamics such as roll pinch-off events, pattern rotation, and the effects of weak 
long-range contributions such as the mean flow.

\begin{figure}[tbh]
\begin{center}
\includegraphics[width=1.5in]{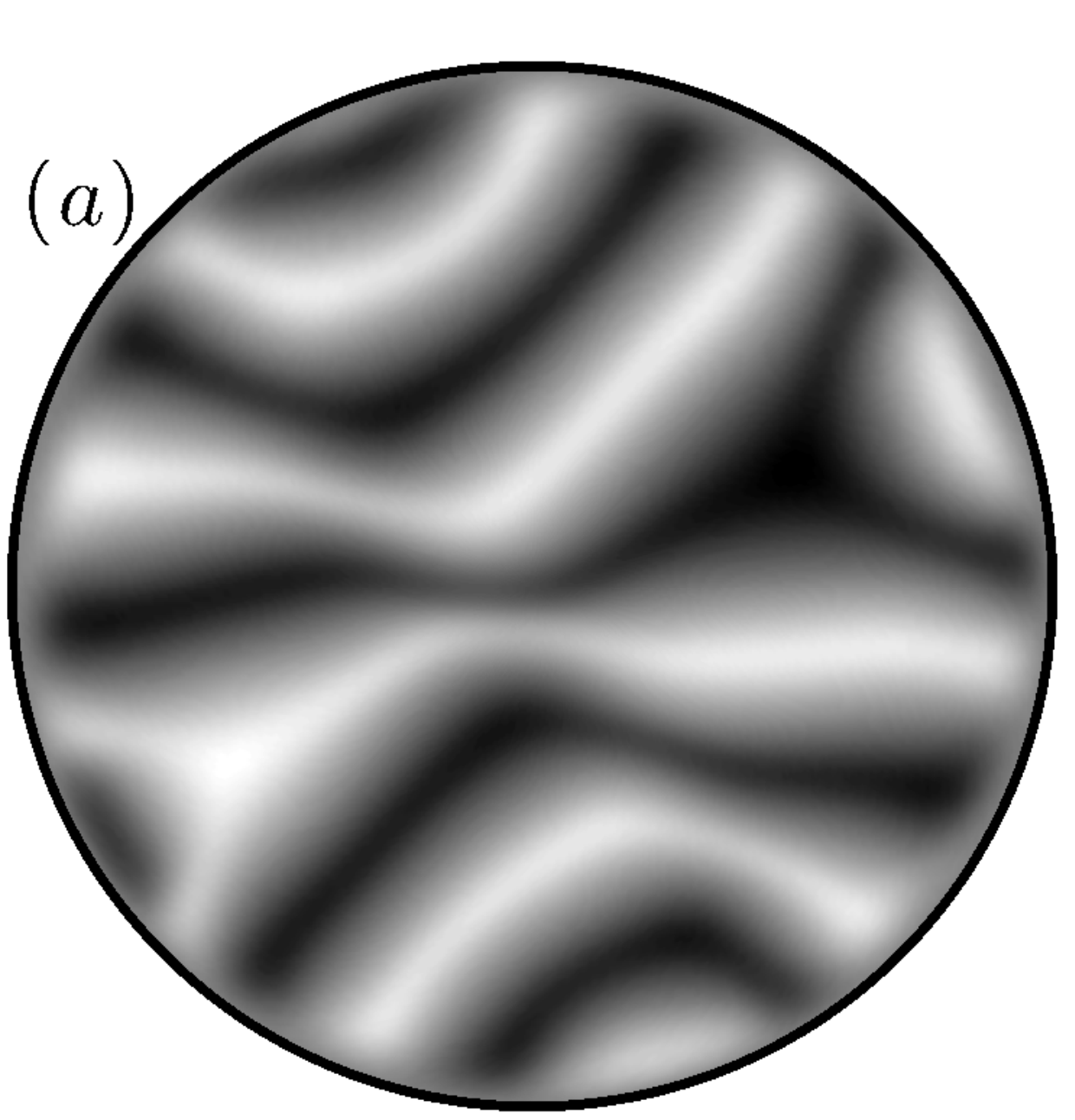} \hspace{0.2cm}
\includegraphics[width=1.5in]{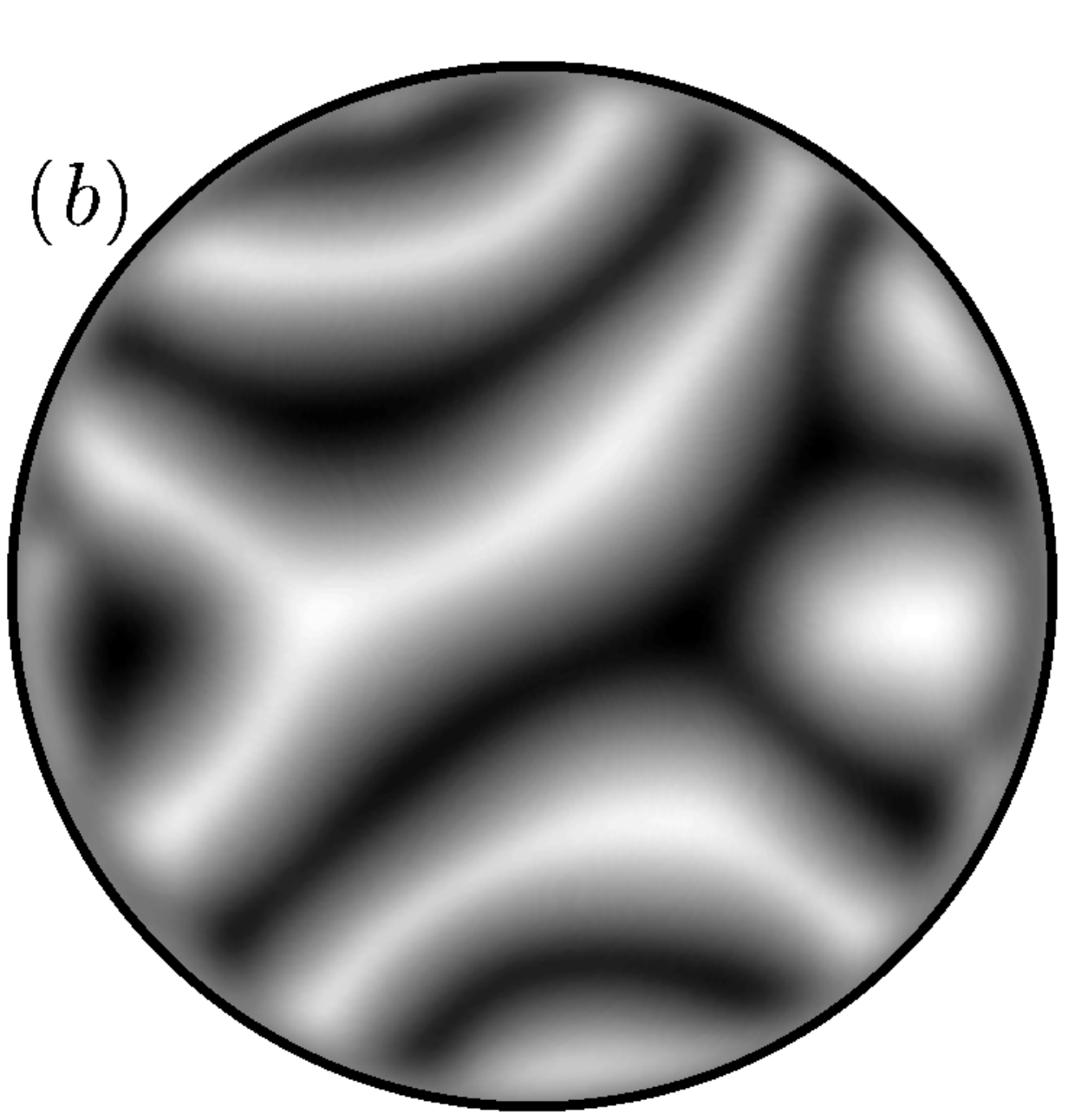} \\ 
\includegraphics[width=3.6in]{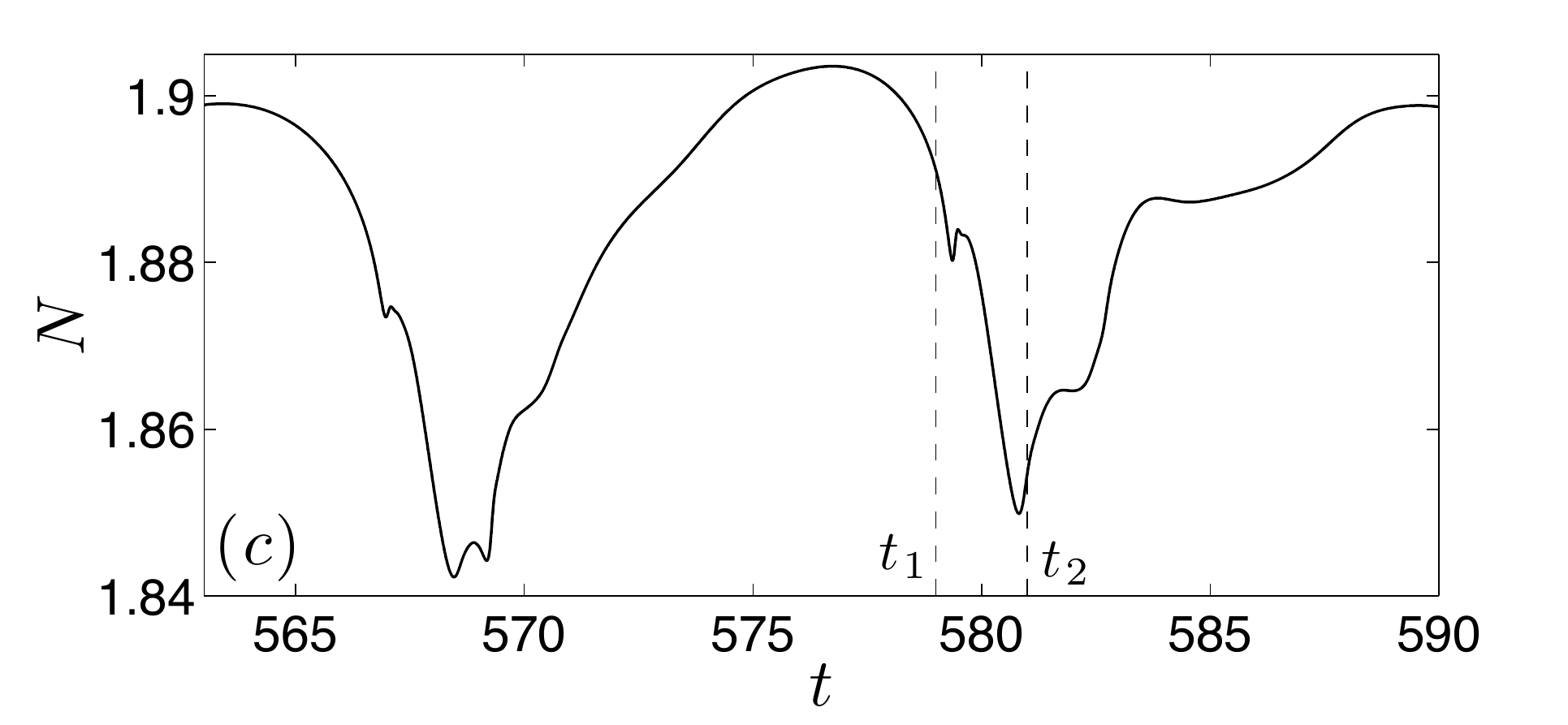}
\end{center}
\caption{The flow field and the variation of the Nusselt number $N$ with time for 
periodic dynamics. The simulation parameters are $\Gamma=5$, $\sigma=1$, 
and $\epsilon=1.93$. (a)~The flow field at $t=579$. (b)~The flow field at $t=581$. (c)~The 
variation $N(t)$ for one period of the dynamics. The vertical lines represent the instances of 
time of the two flow field images.}
\label{fig:g5}
\end{figure}

As either $R$ or $\Gamma$ increase the patterns become very complex making 
it difficult to disentangle distinct features in the pattern dynamics that 
correspond to the variation in the magnitude of the Lyapunov exponents. In 
light of this, we first explore a small cylindrical domain that exhibits periodic 
dynamics in time. The specific parameters used are $\Gamma=5$, $\sigma=1$, and 
$\epsilon=1.93$.  Flow field images are shown in Fig.~\ref{fig:g5}(a)-(b) and the 
variation of the Nusselt number $N$ is shown in Fig.~\ref{fig:g5}(c).

Although $N$ is a global measure of the heat transport through the convection layer 
its variation with time directly corresponds with the topological features 
of the pattern dynamics (c.f.~\cite{paul:2001}).  Figure~\ref{fig:g5}(c) shows one period of the dynamics 
which occurs over a time of $t\approx27$ time units.  The vertical dashed 
lines of Fig.~\ref{fig:g5}(c) indicate the times at which the flow fields in 
Fig.~\ref{fig:g5}(a) and~\ref{fig:g5}(b) are shown.  The dips in $N(t)$ occur 
during roll pinch-off events and the positive spikes occur during dislocation 
annihilation events. Physically, this reflects that the heat transport through 
the convection layer is less efficient in the presence of the defects. The 
remaining smooth features of $N(t)$ correspond to climbing and gliding dynamics.

\begin{figure}[tbh]
\begin{center}
\includegraphics[width=3.25in]{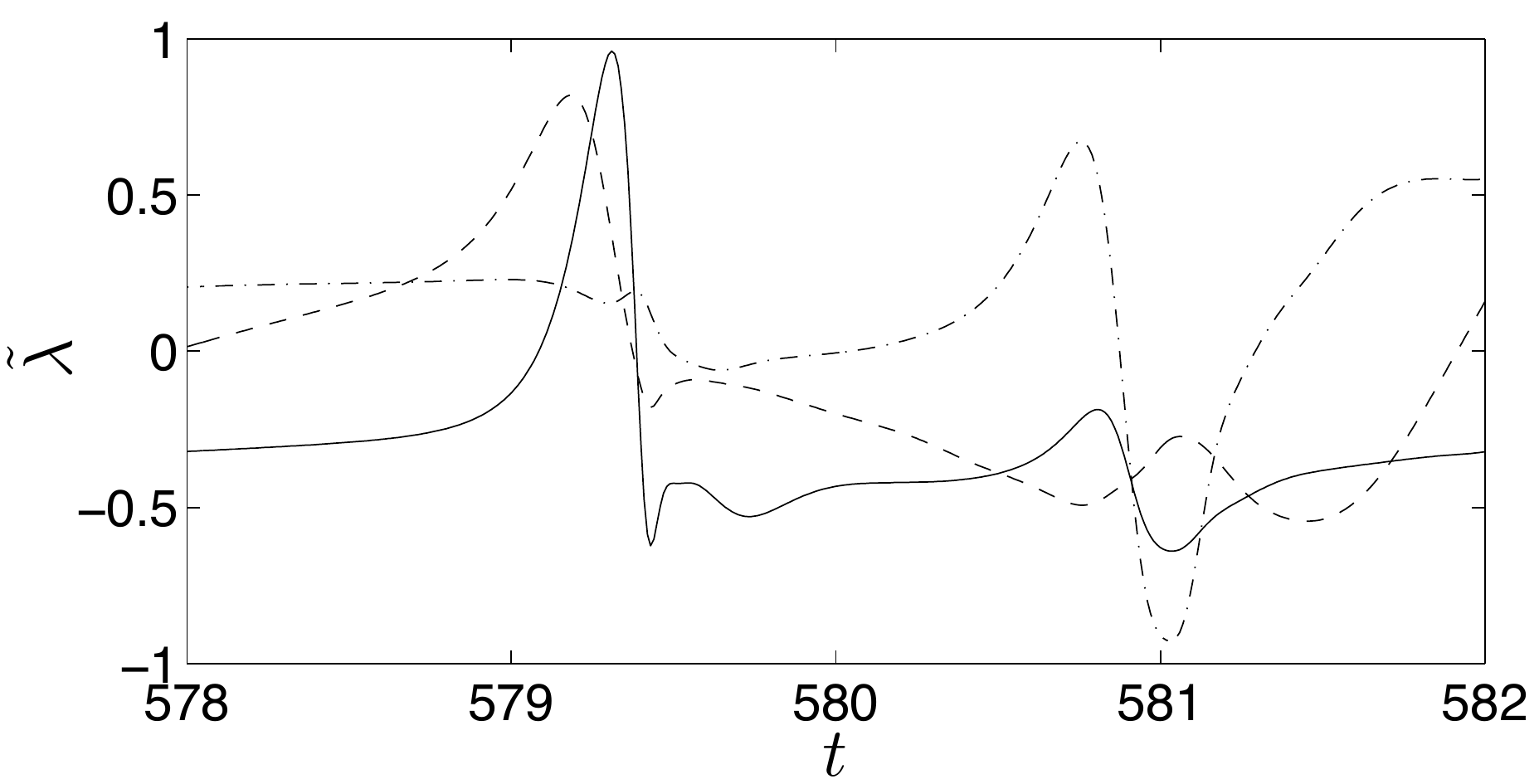}
\end{center}
\caption{The time variation of the first three instantaneous Lyapunov exponents $\tilde{\lambda}$ 
for time periodic dynamics. The simulation parameters are $\Gamma=5$, $\sigma=1$, 
and $\epsilon = 1.93$. The Lyapunov exponents have been normalized by the maximum value 
of $\lambda_1$ for ease of comparison.  The values for $\tilde{\lambda}_1$, $\tilde{\lambda}_2$, 
and $\tilde{\lambda}_3$ are given by the solid, dashed, and dash-dot lines, respectively.}
\label{fig:g5_lyap}
\end{figure}

The time variation of the three largest Lyapunov exponents are shown in 
Fig.~\ref{fig:g5_lyap}.  The exponents have been normalized by the 
maximum value of $\lambda_1$ over this time window in order to compare 
them on a single plot. The normalized exponents are denoted by $\tilde{\lambda}$. 
As expected, the leading order Lyapunov exponent exhibits significant variations at 
the roll pinch-off and annihilation events. The dynamics of the second and 
third exponents tend to follow with some interesting variations.

For example, a closer inspection of the time dynamics near $t\approx579$ reveals 
that the dynamics of $\lambda_1$ correspond precisely with the dynamical 
events of the pattern.  However, the first peak of $\lambda_2$ is before the 
occurrence of the roll annihilation and anticipates this feature.  In addition, 
the roll pinch-off event that occurs at $t\approx581$ results in a peak in 
$\lambda_2$  while both $\lambda_1$ and $\lambda_3$ exhibit dips. The 
dynamics of $\lambda_3$ is much more sensitive to the event that occurs 
near $t\approx581$ than the event near $t\approx579$. 

In order to explore this further for chaotic dynamics we performed a number 
of simulations for a larger cylindrical domain with increased values of the Rayleigh 
number. The specific parameters we used were $\Gamma=10$, $\sigma=1$ 
and $2.51 \le \epsilon \le 4.27$.  An example flow field is shown in 
Fig.~\ref{fig:r9000} for $\epsilon=4.27$.  The dynamics of these patterns 
are quite complex making it very difficult to relate features of the flow field 
dynamics with the variation in the Lyapunov exponents.  In this regime 
there are typically multiple roll pinch-off events occurring simultaneously. 

\begin{figure}[tbh]
\begin{center}
\includegraphics[width=3.0in]{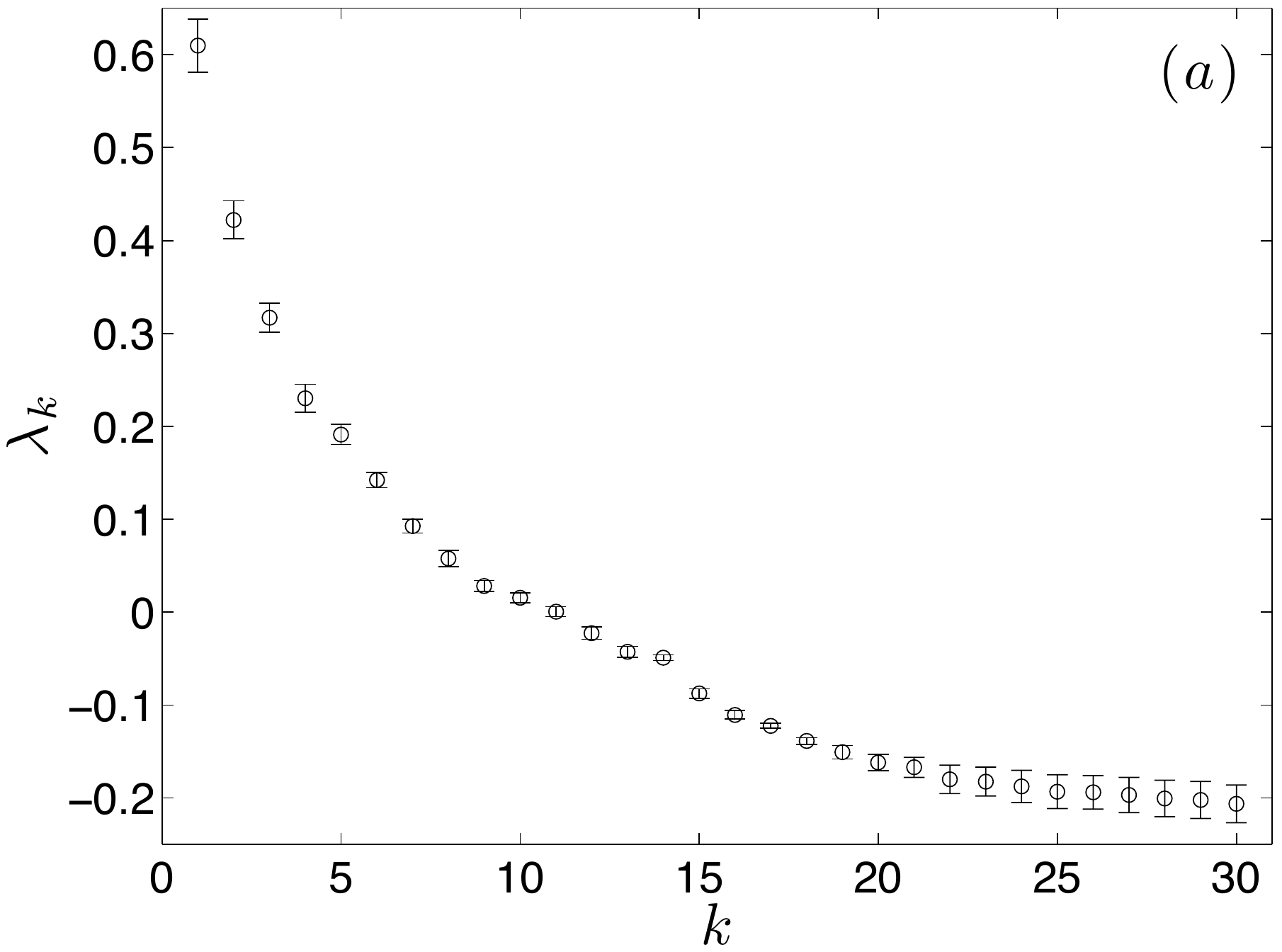}
\includegraphics[width=3.0in]{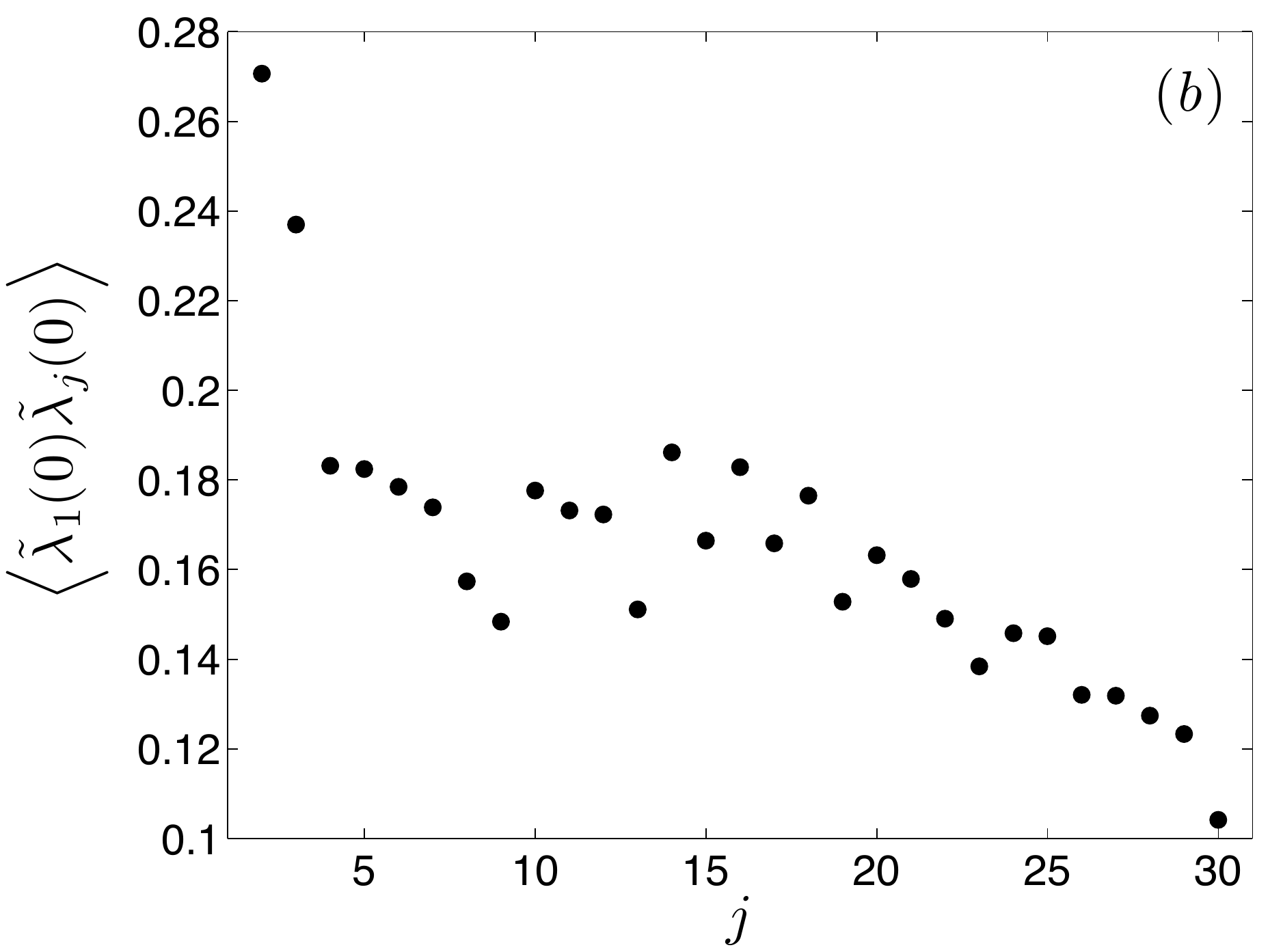} \\
\end{center}
\caption{(a)~The spectrum of Lyapunov exponents $\lambda_k$. Also shown are the 
error bars that are computed from the standard deviation of $\lambda_k$ about its mean 
value at long times. (b)~The instantaneous cross-correlation between the leading order 
Lyapunov exponent and the remaining exponents in the spectra $\lambda_j$ for 
$j= 2 \ldots 30$.  The simulation parameters for both panels are $\sigma=1$, 
$\Gamma=10$, and $\epsilon=2.51$.} 
\label{fig:corr}
\end{figure}

In Fig.~\ref{fig:corr}(a) we plot the spectrum of Lyapunov exponents $\lambda_k$ 
for a convection domain where $\Gamma=10, \sigma=1, \epsilon=2.51$. The 
dynamics is chaotic ($\lambda_1 > 0$) and the error bars represent the 
standard deviation of $\lambda_k$ about its mean value at long times.

Figure~\ref{fig:corr}(b) shows the zero-time cross-correlation between $\tilde{\lambda}_1$ 
and $\tilde{\lambda}_j$ where $j=2, \ldots, N_\lambda$, we have first subtracted off the 
mean value of each of the Lyapunov exponents, and $N_\lambda$ is the number of Lyapunov 
exponents computed for that value of $\epsilon$.  We find a positive cross-correlation 
for all of the exponents $\tilde{\lambda}_j$.  The first several exponents have the largest 
cross-correlation with $\tilde{\lambda}_1$ which is then followed by a rather 
uniform fall-off with increasing $j$.  These results suggest that all of the 
exponents tend to exhibit variations together.  In these patterns the dynamics 
are dominated by roll pinch-off events suggesting that all of the exponents 
are sensitive to these events.


\subsection{The Dynamics of the Leading Order Lyapunov Vector}

The spatial and temporal dynamics of the leading order Lyapunov vector provides insight 
into regions of the flow field experiencing the largest growth in the perturbation equations. 
This has been used to identify non-repeating roll pinch-off events as significant contributors 
to the overall disorder in a chaotic convection flow field~\cite{scheel:2006}.  It has also been shown that 
Rayleigh-B\'{e}nard convection exhibits extensive chaos for finite cylindrical geometries 
using systems parameters that yield spiral defect chaos. For the parameters used 
by Paul \textit{et al.}, $\epsilon = 2.51$ and $\sigma=1$, the onset of extensivity 
occurred for a system size of $\Gamma \approx 7$~\cite{paul:2007}.  It is expected that 
extensive chaos occurs for convection layers that have reached a large-system limit 
where the influence of the lateral sidewalls have become reduced.

\begin{figure}[tbh]
\begin{tabular}{cc}
\includegraphics[width=2.5in]{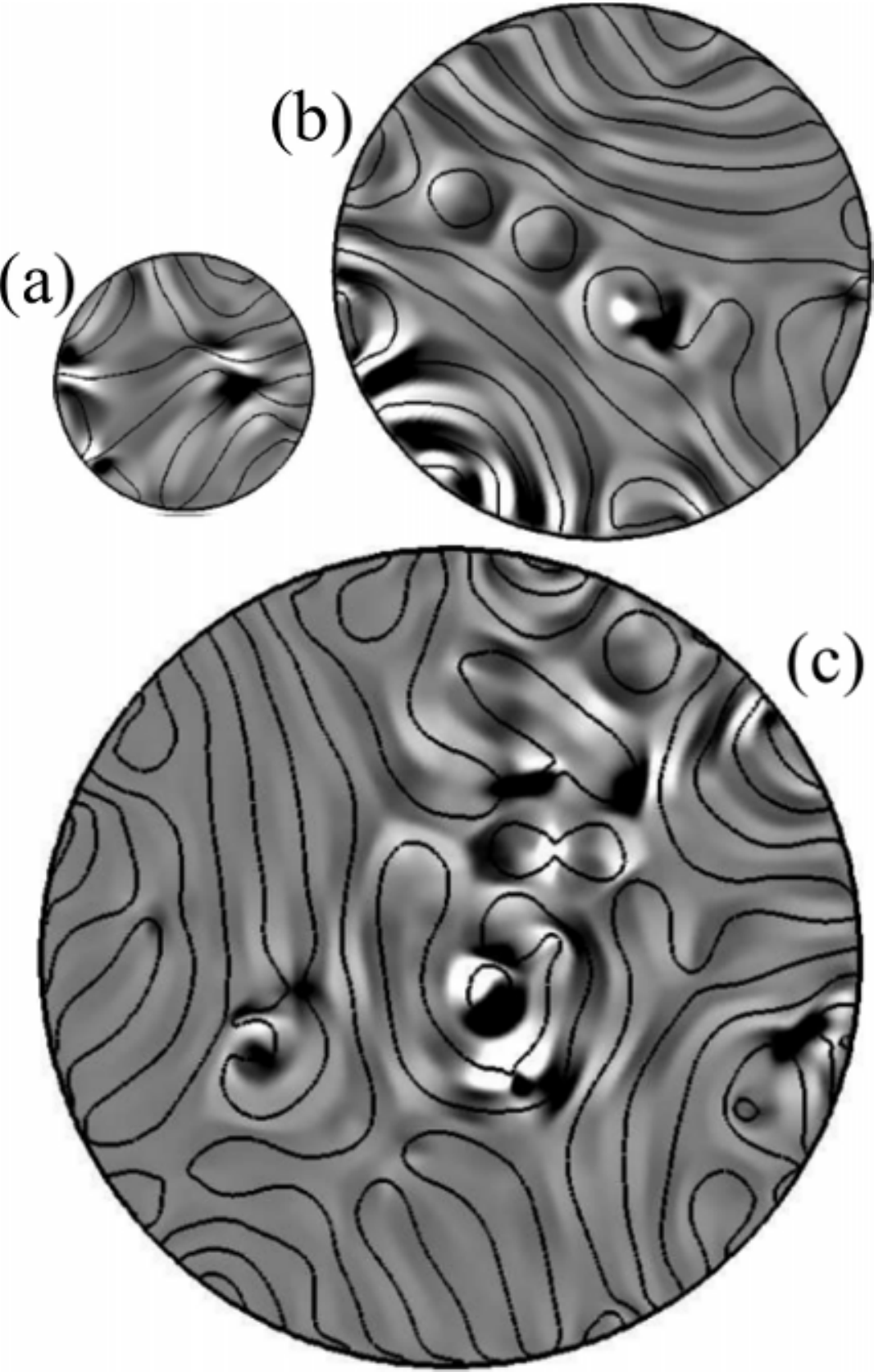} \\
\includegraphics[width=2.5in]{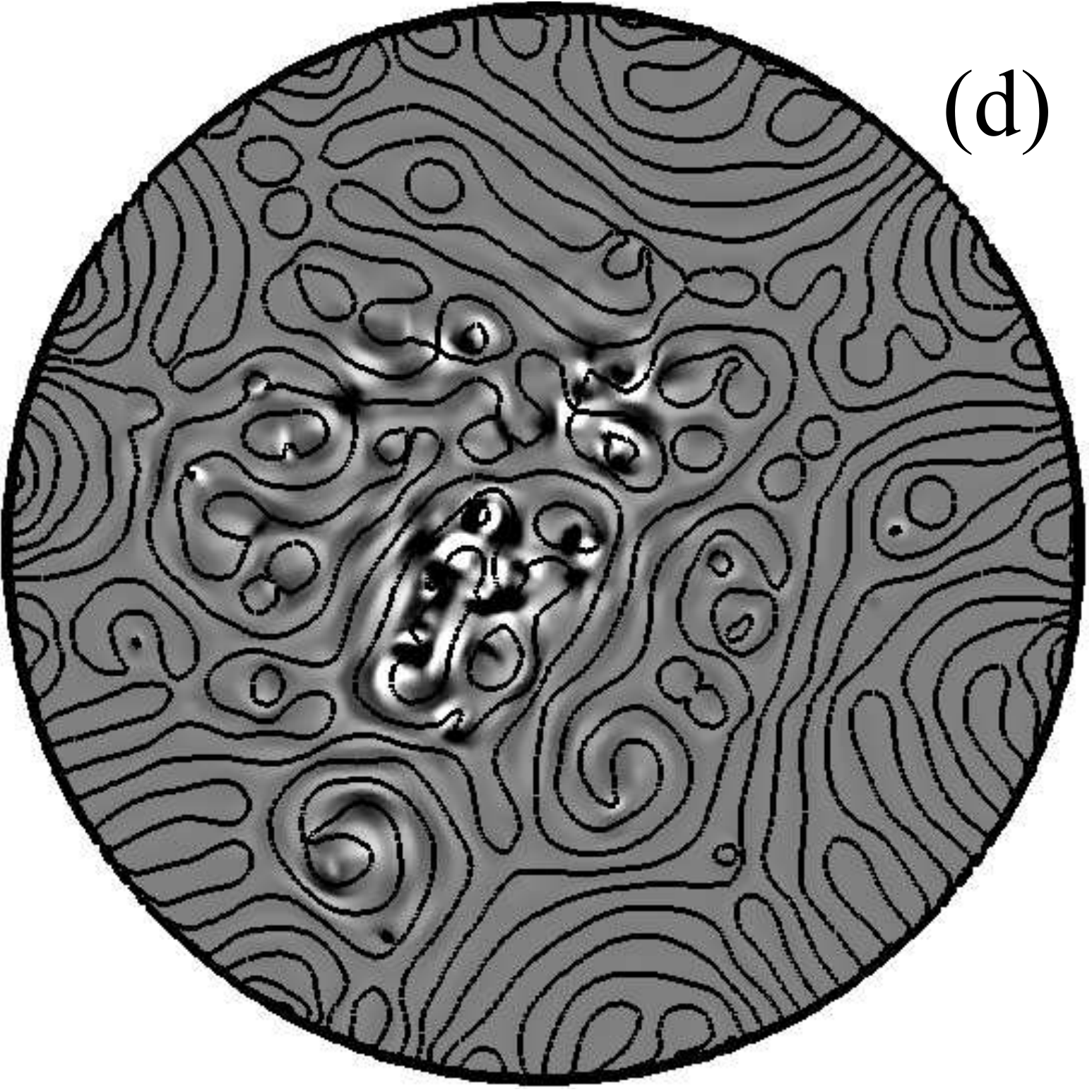} 
\end{tabular}
\caption{Overlay of a grey-scale contours of the midplane temperature perturbation 
field with solid black lines representing the convection roll boundaries for different 
aspect ratios: (a) $\Gamma=5$, (b) $\Gamma=10$, (c) $\Gamma=15$, and 
(d) $\Gamma=30$. For $\Gamma=30$ the image is plotted at half scale to be 
able to fit on this figure. The parameters are $\epsilon=2.51$ and $\sigma=1$.} 
\label{fig:vectors}
\end{figure}

In order to explore this further we have performed very long-time numerical simulations 
for cylindrical geometries over a range of aspect ratios $5 \le \Gamma \le 30$ where 
$\epsilon = 2.51$ and $\sigma=1$.  In these simulations we have computed the 
leading-order Lyapunov exponent and Lyapunov vector. Figure~\ref{fig:vectors} 
shows grey-scale contours of the leading order Lyapunov vector overlaid with solid black lines indicating the 
convective roll pattern.  The Lyapunov vector is plotted using the value of the thermal 
perturbation field at the horizontal mid-plane. In this figure light regions indicate large positive 
values, dark regions indicate large negative values, and grey regions represent regions of small 
growth.  The images of Fig.~\ref{fig:vectors} suggest that the smaller domains are 
dominated by large values of the perturbation at the lateral boundaries. This transitions 
to dynamics with large perturbations in the bulk of the domain away from the 
sidewalls for the larger aspect ratio systems. The location of occurrence of 
the largest perturbations also shows a transition. In small domains, mostly bending rolls 
cause large perturbations; but in large domains, they are associated with the 
dislocation defects initiated by roll pinch-off events.

\begin{figure}[!Htb]
\begin{tabular}{cc}
\includegraphics[width=2.5in]{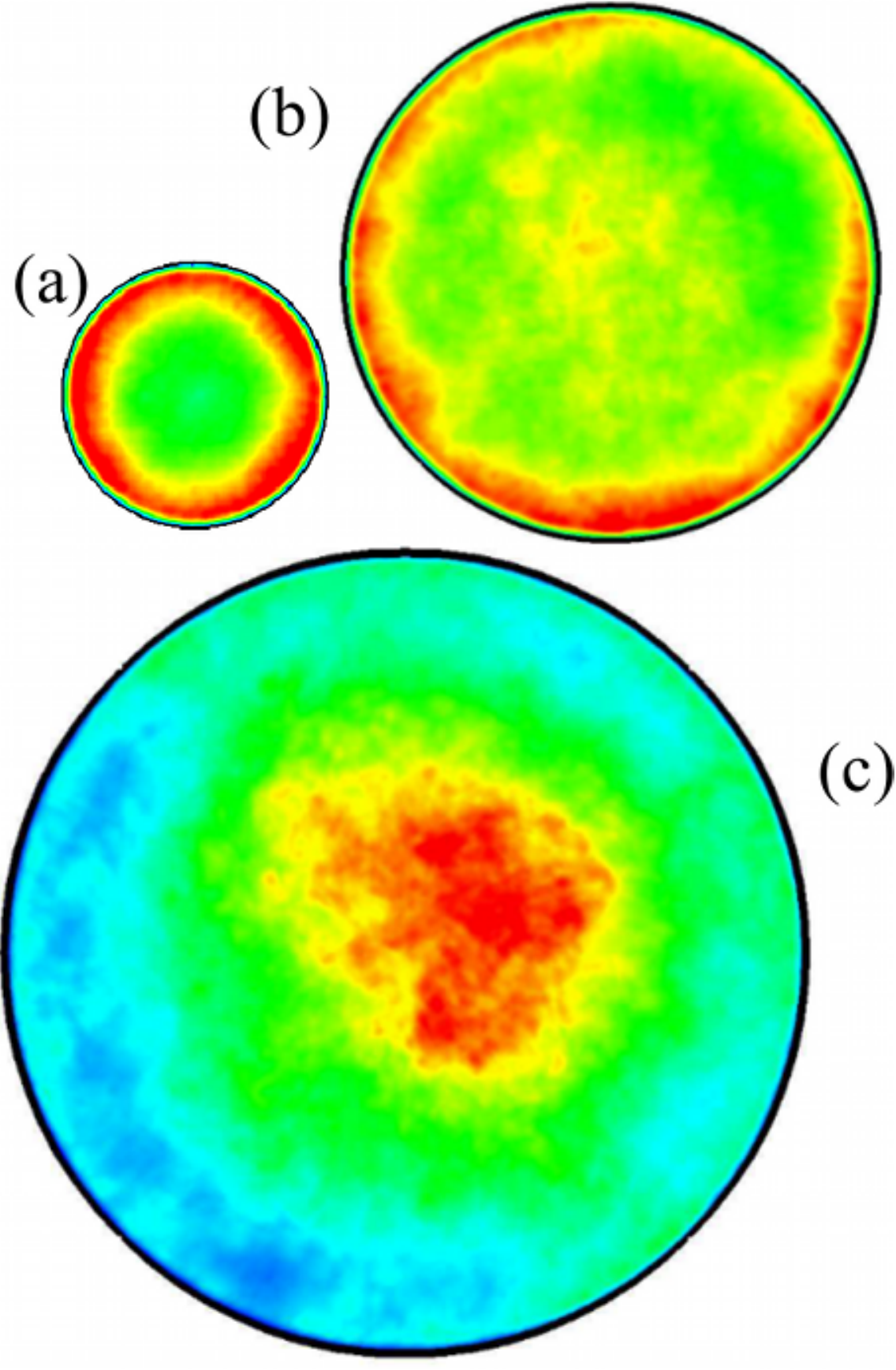} \\
\includegraphics[width=2.5in]{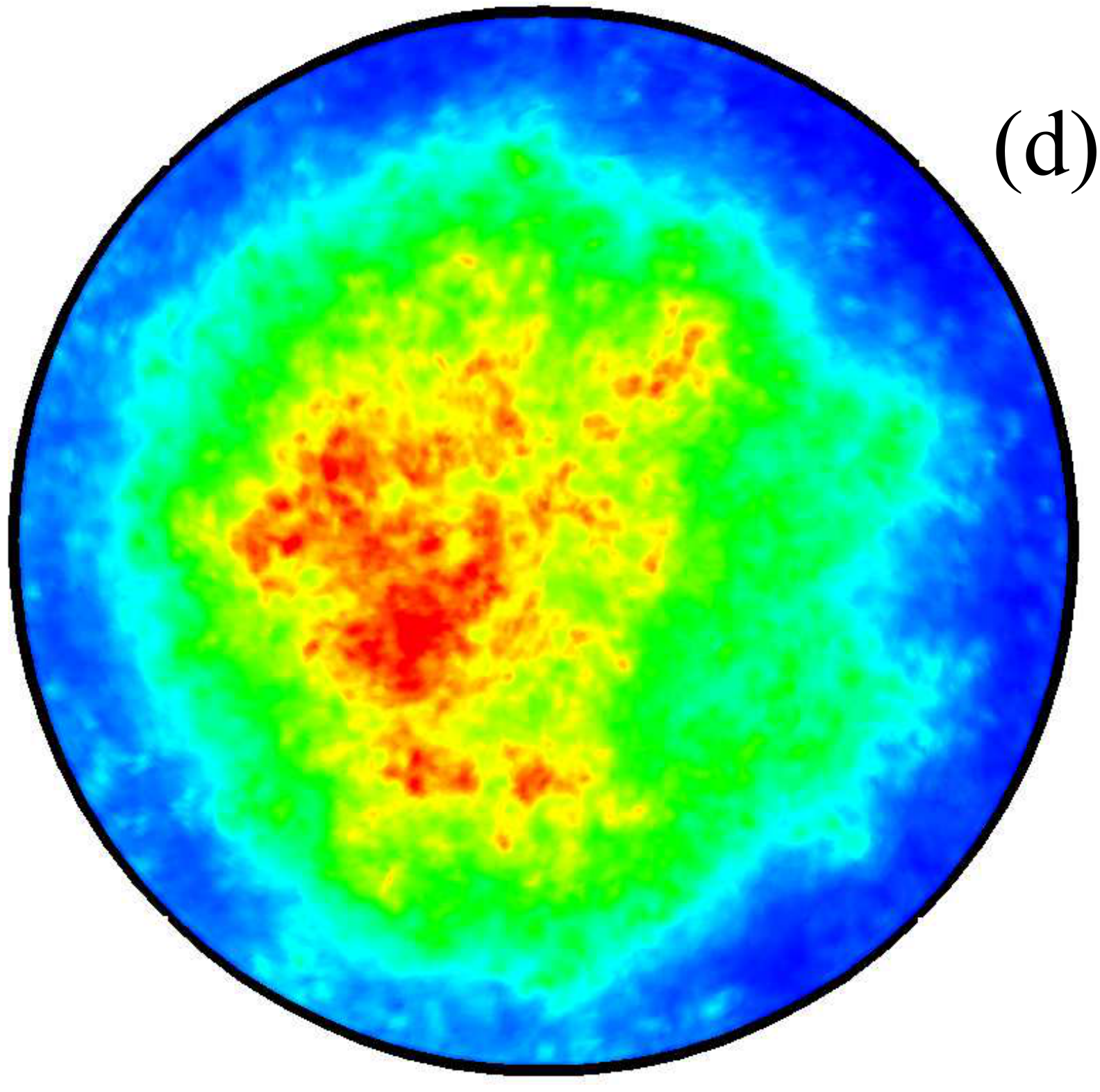}  
\end{tabular}
\caption{(Color online) The spatial variation of the time-averaged magnitude of the thermal 
perturbation field $\left< \delta T(x,y) \right>_t$ evaluated at the horizontal mid-plane. 
(a)~$\Gamma=5$, (b)~$\Gamma=10$, (c)~$\Gamma=15$, and (d)~$\Gamma=30$. 
The simulation parameters are  $\epsilon=2.51$ and $\sigma=1$. The image 
for $\Gamma=30$ is plotted at half scale to fit on this figure. In the color contour, red regions (located primarily near the boundary for small $\Gamma$ and at the bulk of the domain for large $\Gamma$) correspond to the large magnitude of the perturbation and blue regions (located mainly at the bulk of the domain for small $\Gamma$ and near the boundary for large $\Gamma$) associate with the small magnitude of the perturbation.}
\label{fig:lyap_average}
\end{figure}

\begin{figure}[!Htb]
\begin{center}
\includegraphics[width=3.0in]{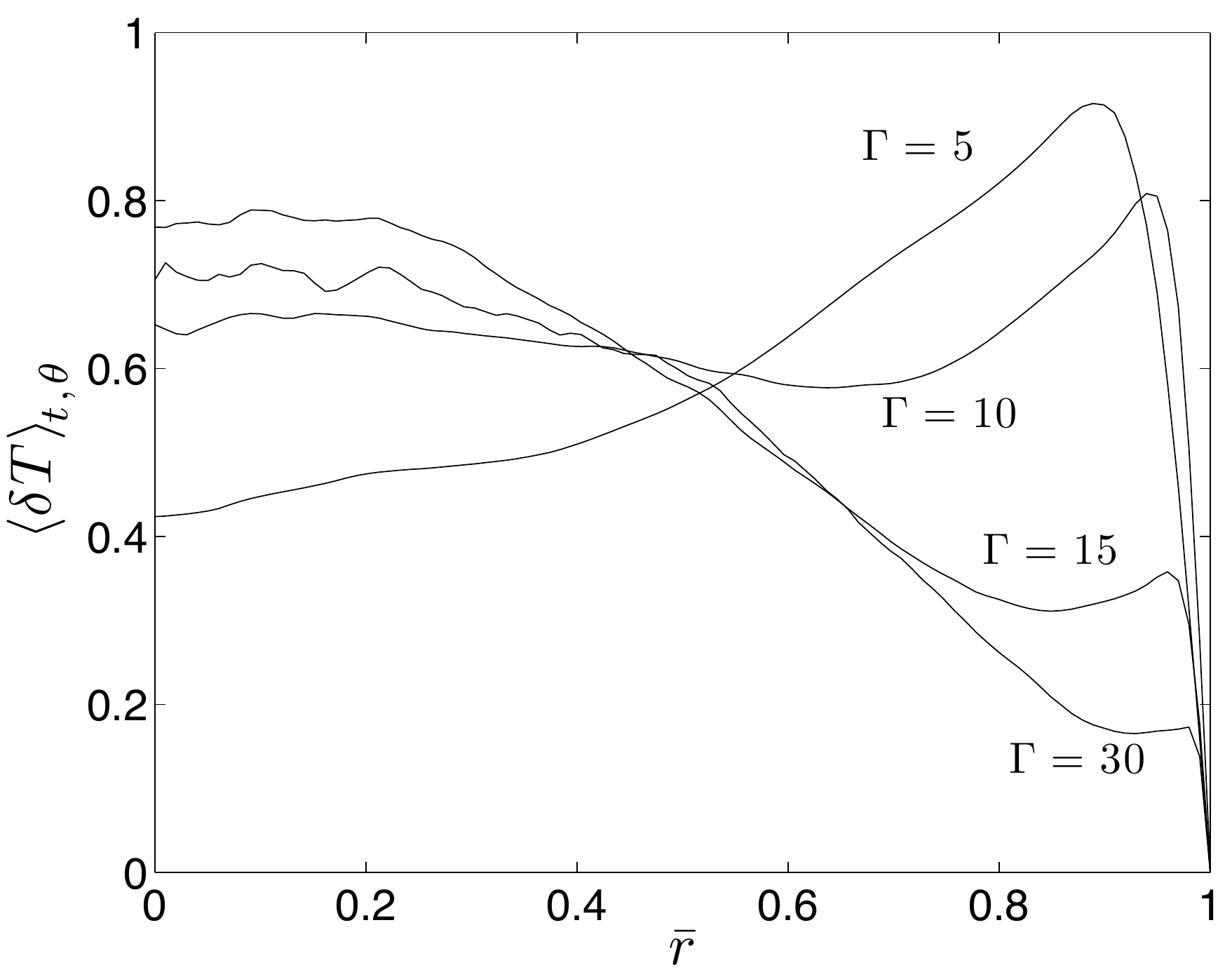}
\end{center}
\caption{The radial variation of the azimuthal and time averaged thermal perturbation 
field $\left< \delta T(\bar{r}) \right>_{t,\theta}$. The aspect ratios are $5 \le \Gamma \le 30$.} 
\label{fig:average}
\end{figure}

In order to investigate this further we have computed the time average of the magnitude of 
the leading order Lyapunov vector given by,
\begin{equation}
	\left< \delta T(x,y) \right>_t = \frac{1}{N_s}\sum_{i=1}^{N_s} |\delta T^{(1)}(x,y,z=0.5,t_i)|,
\label{eq:pert1}
\end{equation}
where $t_i$ is the time of the corresponding perturbation field, $N_s \approx 10^3$ is 
the total number of perturbation fields, and the notation $\left< \cdot \right>_t$ is used to 
indicate the time-average. The spatial distribution of the time-averaged perturbation 
fields are shown in Fig.~\ref{fig:lyap_average}.  In Fig.~\ref{fig:lyap_average} 
red indicates regions of large values of the magnitude (located primarily near the boundary for small domains and at the bulk of the domain for large domains) and blue represents regions of small values of the 
magnitude (located mainly at the bulk of the domain for small domains and near the boundary for large domains). The asymmetry in the azimuthal direction of the averaged perturbation 
fields is most likely a result of the finite time of the simulations and the particular choice 
of random initial conditions.  These simulations are quite computationally expensive 
and we have not explored this aspect further.  In order to explore the variation with 
the radial coordinate we have computed the azimuthal average of the time-averaged 
perturbation fields using, 
\begin{equation}
	\left< \delta T(\bar{r}) \right>_{t,\theta} = \frac{1}{N_{\theta}} \sum_{i=1}^{N_\theta} \left< \delta T(\bar{r},\theta) \right>_t,
\label{eq:pert2}
\end{equation}
where the notation $\left< \cdot \right>_{t,\theta}$ indicates time and azimuthal averaging, $\bar{r}=r/\Gamma$ 
is the normalized radial coordinate whose origin is in the center of the domain, and $N_{\theta}=400$ 
is the number of points used in computing the azimuthal average.  The radial variation of 
$\left< \delta T(\bar{r}) \right>_{t,\theta}$ is plotted in Fig.~\ref{fig:average}.  The transition from 
dynamics with significant perturbations at the boundaries to dynamics with significant 
perturbations away from the walls is evident.

\subsection{The Variation of the Fractal Dimension with System Parameters}

The variation of the fractal dimension with system parameters can provide insights 
into the nature and composition of the underlying high-dimensional attractor describing 
the chaotic dynamics.  For Rayleigh-B\'{e}nard convection 
$D_\lambda = D_\lambda(\epsilon,\sigma,\Gamma)$.  Our approach is to compute the 
variation of $D_\lambda$ with one of the parameters while the remaining two are 
held constant.  This has allowed us to quantitatively probe the underlying attractor for 
three different limiting cases. By increasing the system size while holding $\epsilon$ and 
$\sigma$ constant we are able to quantify the increase in the fractal dimension in the 
spatiotemporal chaos limit~\cite{cross:1993}.  When the driving $\epsilon$ is increased 
while holding $\Gamma$ and $\sigma$ constant we are able to quantify the 
increase in the fractal dimension with the addition of new degrees of freedom 
as the system approaches the strong driving limit. Lastly, the magnitude of the 
Prandtl number $\sigma$ is inversely related to the magnitude of the mean 
flow. By the varying $\sigma$ while holding $\Gamma$ and $\epsilon$ 
constant we quantify the variation of the fractal dimension as the system 
transitions from non-potential to potential dynamics.

The variation of the fractal dimension with system size is expected to be extensive where
\begin{equation}
D_\lambda \propto \Gamma^{d_s}
\end{equation}
in the large system limit and $d_s$ is the number of spatially 
extended directions~\cite{ruelle:1982}.  For Rayleigh-B\'{e}nard convection in large shallow 
layers $d_s=2$.  Extensive chaos has been demonstrated in large periodic convection 
layers~\cite{egolf:2000} and in finite cylindrical convection layers~\cite{paul:2007}.  Deviations 
from extensive chaos for small changes in system size has been proposed as a means to identify 
a length scale associated with the fundamental structures composing spatiotemporal 
chaos~\cite{fishman:2006}. Deviations from extensivity have been found using the complex 
Ginzburg-Landau equation~\cite{fishman:2006}, the Lorenz-96 equations~\cite{karimi:2010} and 
systems of coupled map lattices~\cite{ohern:1996}. However, microextensivity has been found 
for the Kuramoto-Sivashinsky equation~\cite{tajima:2002}.  For Rayleigh-B\'{e}nard convection 
we have found that the slow and noisy convergence of $D_\lambda$ (see Fig.~\ref{fig:convergence}) 
precludes such an investigation using currently available algorithms and computing resources.

\begin{figure}[!Htb]
\begin{center}
\includegraphics[width=3in]{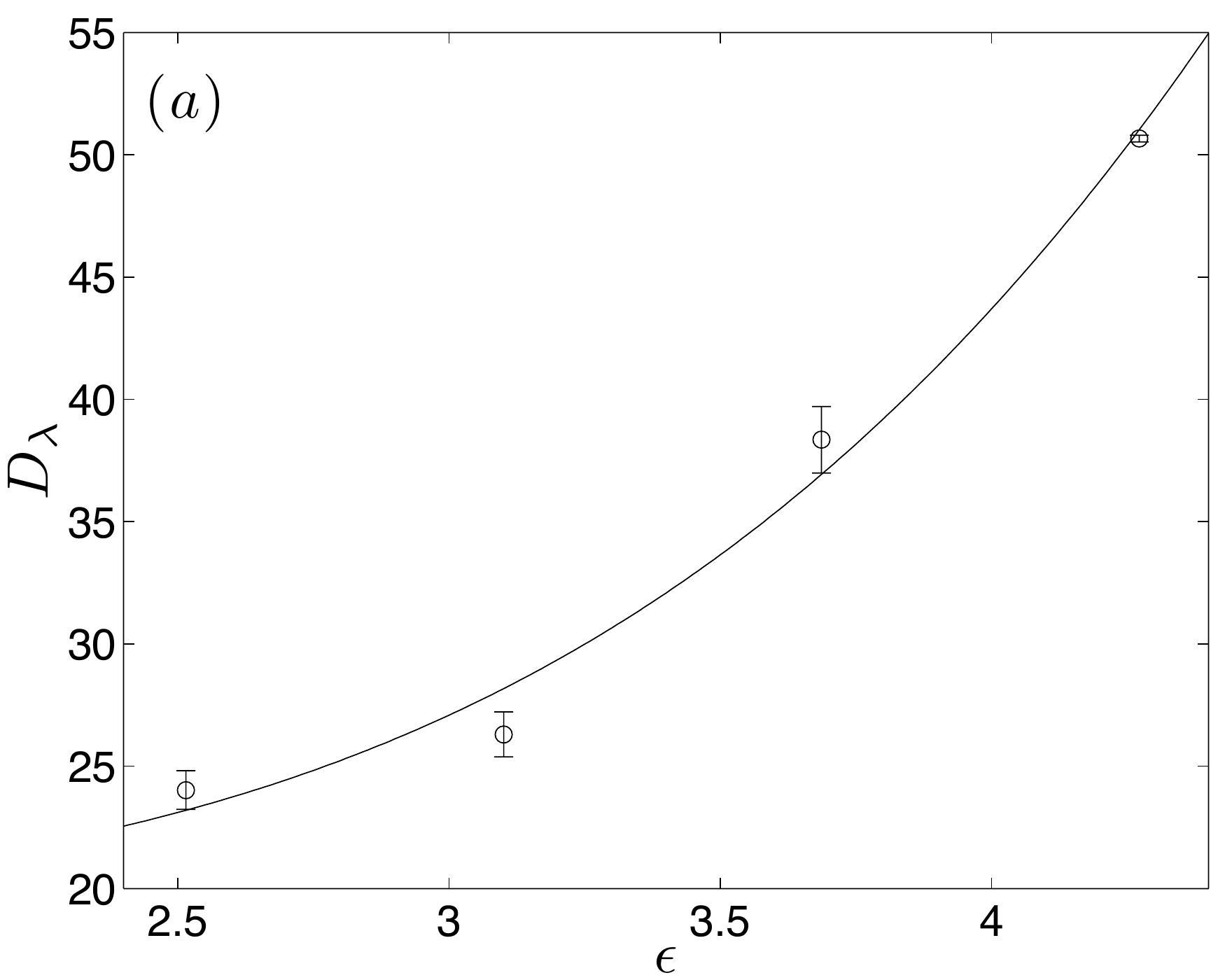}
\includegraphics[width=3in]{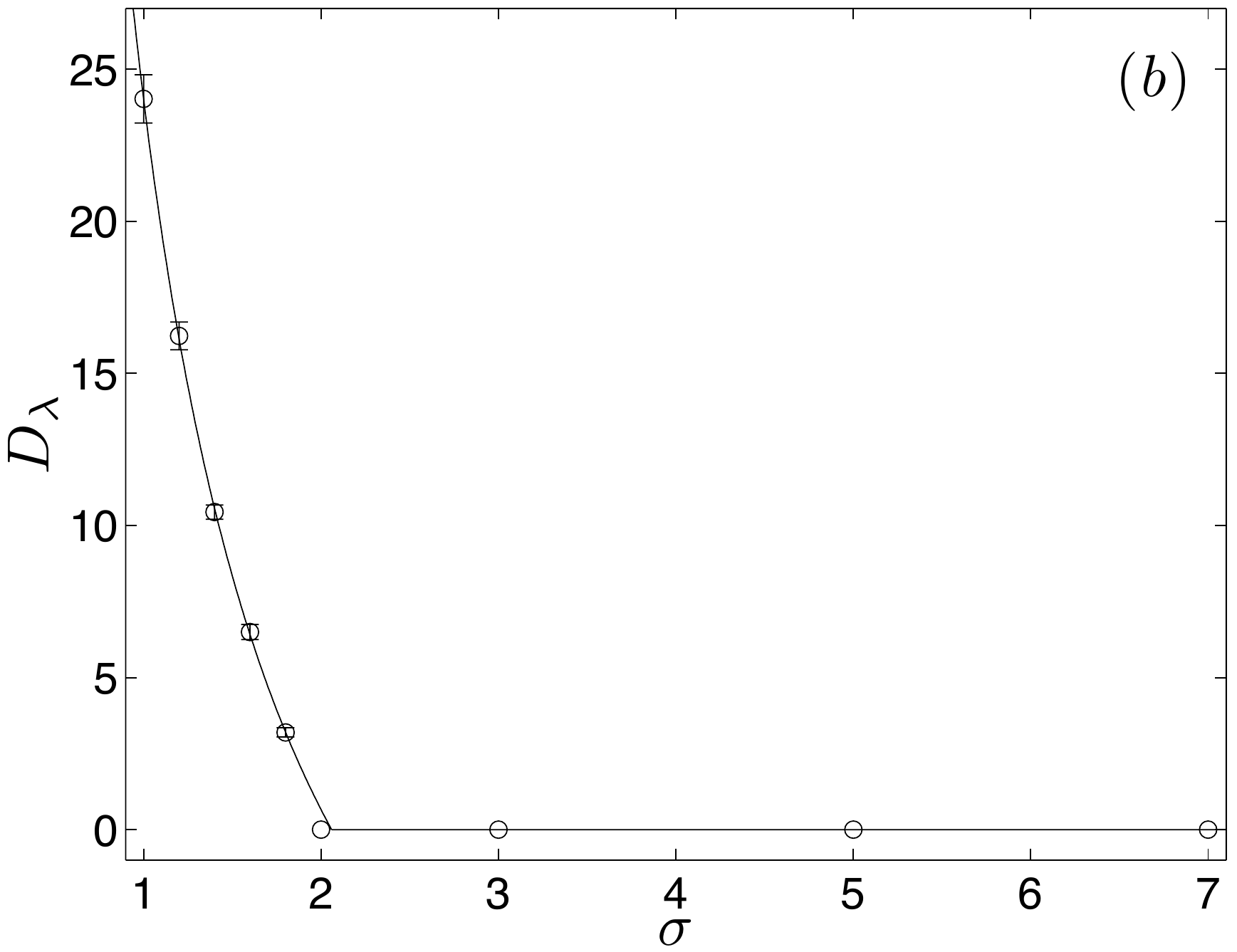}
\end{center}
\caption{(a) The variation of the fractal dimension with Rayleigh number for $\Gamma=10$ and 
$\sigma=1$. The circles are data points from the simulations and the solid line is the curve 
fit $D_\lambda=0.095\epsilon^{4}+19.4$. (b)~The variation of the fractal dimension with 
Prandtl number for $\Gamma=10$ and $\epsilon=2.51$. The circles are data points from the 
simulations and the solid line is a power-law curve fit as $D_\lambda=44.95\sigma^{-1.06}-20.91$ for $1\leqslant \sigma < 2$ and $D_\lambda=0$ for $\sigma \geqslant 2$.} 
\label{fig:dimension}
\end{figure}

The variation of $D_\lambda$ with $\epsilon$ and $\sigma$ is shown in Fig.~\ref{fig:dimension}. 
Fig.~\ref{fig:dimension}(a) illustrates the variation of $D_\lambda$ with $\epsilon$ where 
$\sigma=1$ and $\Gamma=10$.  A typical flow field pattern for the largest value of the 
forcing $\epsilon=4.27$ is shown in Fig.~\ref{fig:r9000}.  The error bars represent the standard 
deviation of $D_\lambda$ about its mean value in the large-time limit.  The solid line through the 
data is a curve fit given by
\begin{equation}
D_\lambda = \alpha \epsilon ^4 + \beta
\end{equation}
where $\alpha=0.095$ and $\beta=19.4$. This relationship is only useful for $\epsilon \gtrsim 2.5$.  
For smaller value of $\epsilon$ there must be a transition not captured in our data that would 
yield a vanishing value of the fractal dimension at some positive and finite value of $\epsilon$. It is 
possible that our curve fit remains valid for Rayleigh numbers larger than what is shown, however 
without further evidence this remains speculative.  It is interesting to note that Sirovich and 
Deane~\cite{sirovich:1991} found that the fractal dimension increases linearly with Rayleigh 
number from numerical simulations of turbulent Rayleigh-B\'enard convection ($\epsilon \approx 70$) 
in a small periodic box with free-slip boundaries. 

\begin{figure}[!Htb]
\begin{center}
\includegraphics[width=3in]{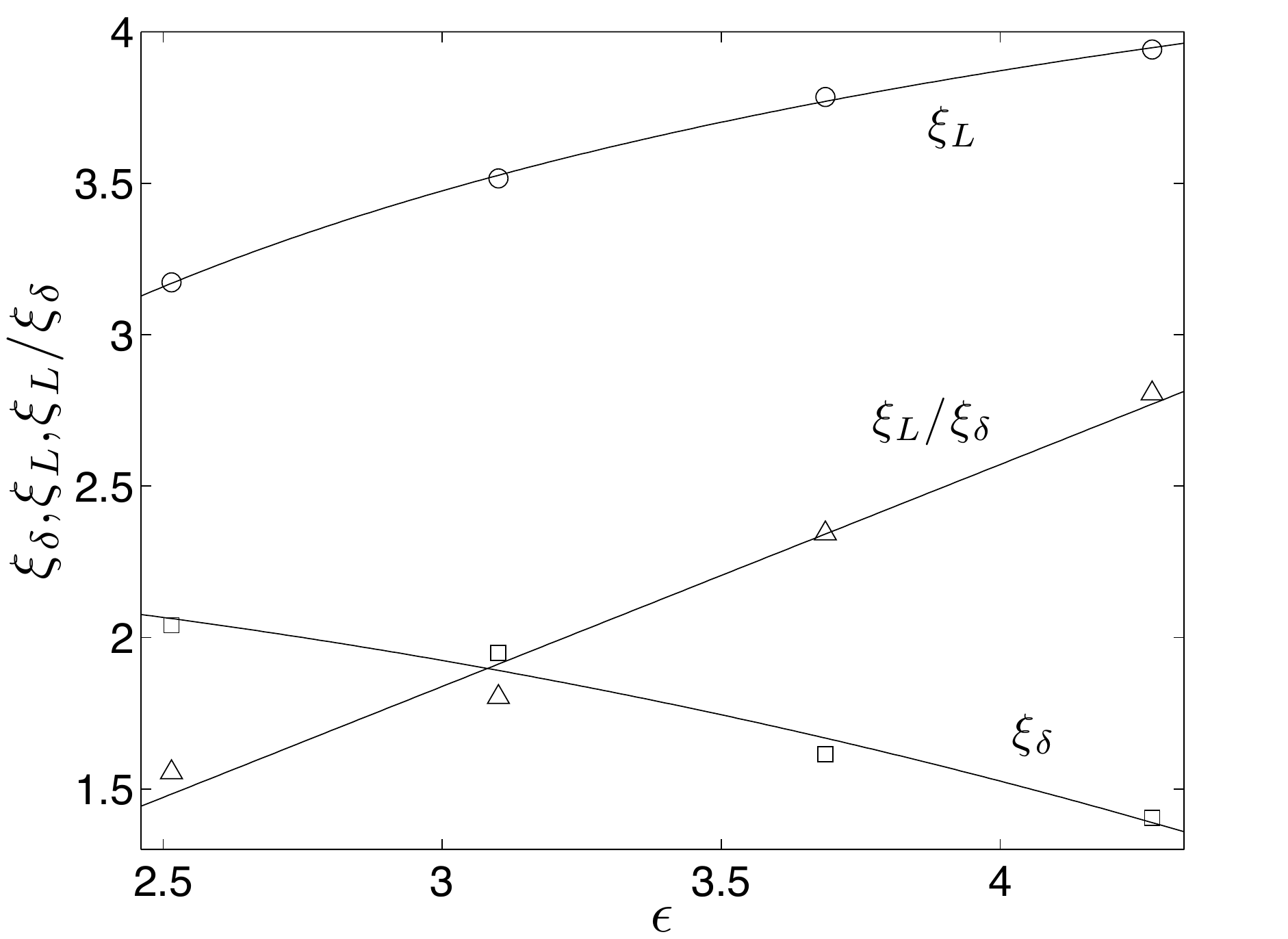}
\end{center}
\caption{The variation of the natural chaotic length scale ($\xi_\delta$), the wavelength 
of the pattern ($\xi_L$), and the ratio of $\xi_L/\xi_\delta$ with $\epsilon$ for $\Gamma=10$ 
and $\sigma=1$. The open squares show $\xi_\delta$, the open circles show $\xi_L$, 
the open triangles demonstrate $\xi_L/\xi_\delta$, and the solid lines illustrate curve fits 
for $\xi_\delta$ and $\xi_L$ as $\xi_\delta=2.32-0.03 \epsilon^{2.4}$ and 
$\xi_L=5.07-4.75 \epsilon^{-0.99}$ and linear fit for the ratio as $\xi_L/\xi_\delta=0.73 \epsilon-0.36$.} 
\label{fig:length}
\end{figure}

The fractal dimension can be used to provide an estimate for a natural chaotic 
length scale~\cite{cross:1993},
\begin{equation}
\xi_\delta \equiv \left( \frac{D_\lambda}{\Gamma^{d_s}} \right)^{-1/d_s}
\label{eq:xi}
\end{equation}
where a volume of size $\xi_\delta^{d_s}$ contains a single chaotic degree of 
freedom on average.  The variation of $\xi_\delta$ with $\epsilon$ is shown in 
Fig.~\ref{fig:length} where it decreases from approximately 2 to 1.5 over the 
range of $\epsilon$ explored.  In order to compare this with features of the 
spatial patterns we have computed the time averaged value of the pattern 
wavelength $\xi_L$ from the structure factor~\cite{cross:1993}.  The pattern 
wavelength increases from approximately 3 to 4 over the range explored.  The 
ratio $\xi_L/\xi_\delta$ provides an estimate for the number of chaotic degrees 
of freedom per wavelength of the flow field pattern and is also shown on 
Fig.~\ref{fig:length}.  This indicates that the number of chaotic degrees of 
freedom per wavelength of the pattern is increasing with increasing $\epsilon$. 
This is reflected by the occurrence of smaller scale features in the pattern images.

\begin{figure}[!Htb]
\begin{tabular}{cc}
\includegraphics[width=1.5in]{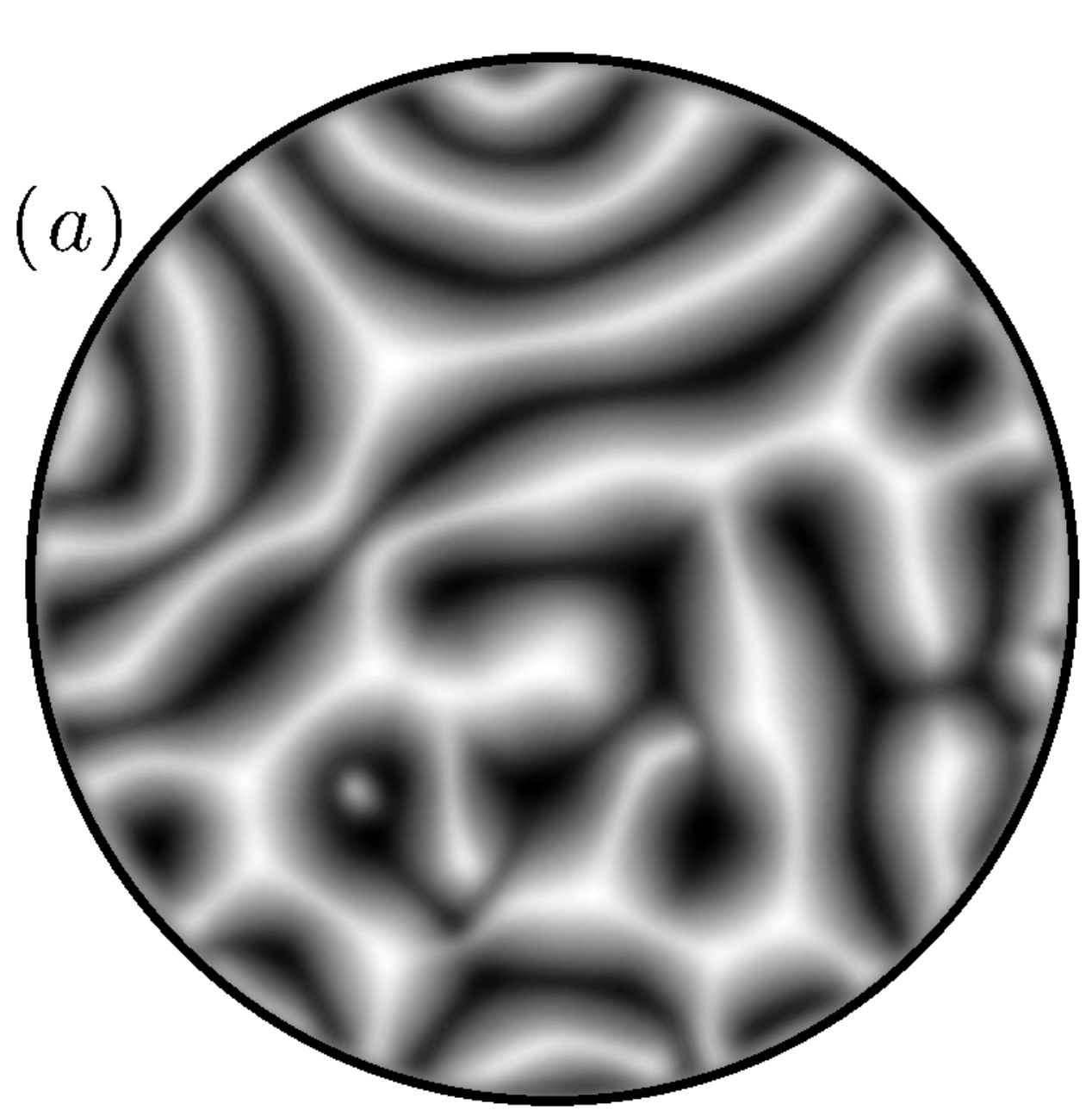}
\includegraphics[width=1.5in]{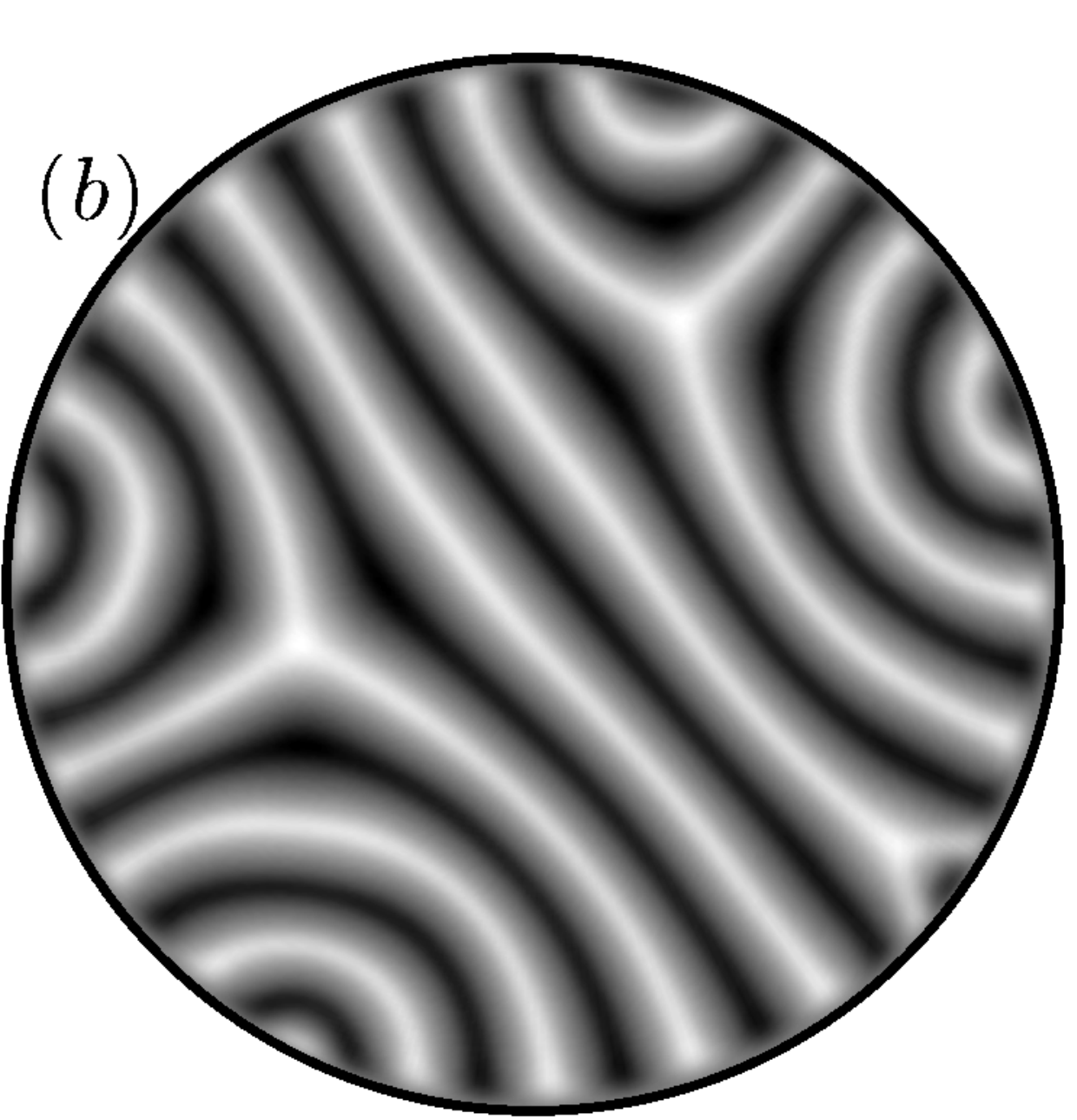} \\ 
\includegraphics[width=1.5in]{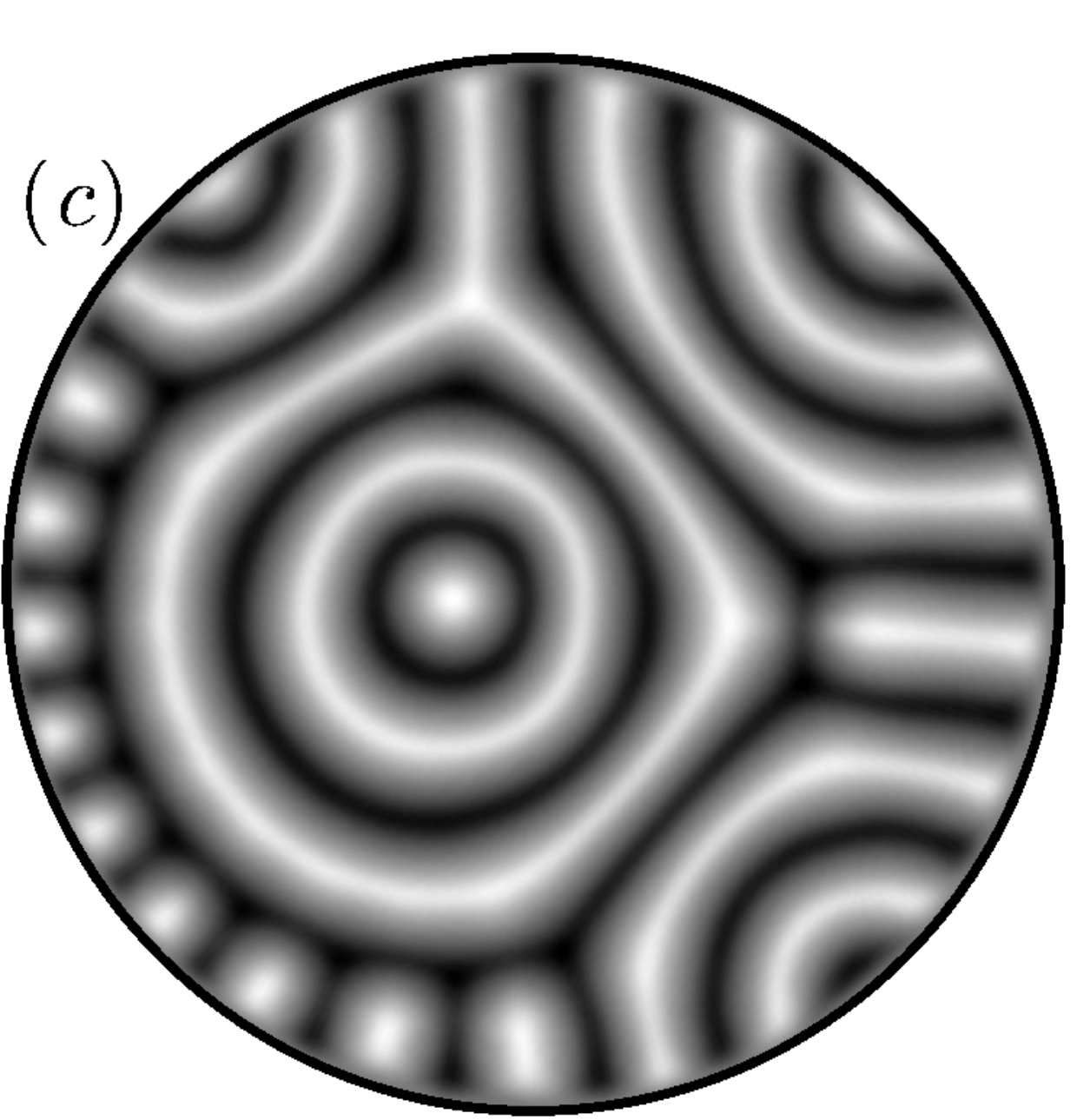}
\includegraphics[width=1.5in]{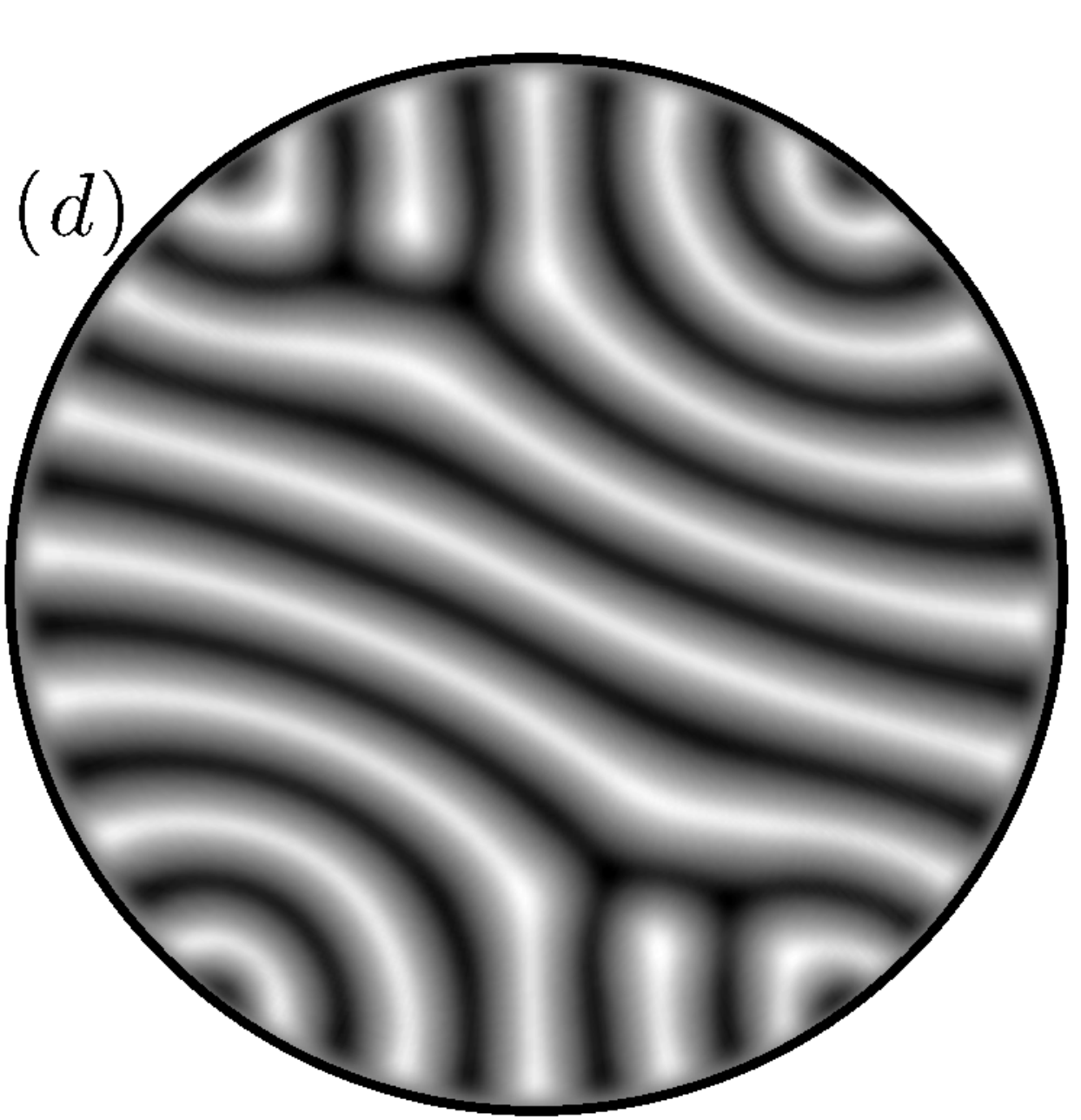}
\end{tabular}
\caption{The flow field patterns for different Prandtl numbers. In each case $\epsilon=2.51$ and 
$\Gamma=10$. Panel (a)~$\sigma = 1$, (b)~$\sigma = 3$, 
(c)~$\sigma = 5$, (d)~$\sigma = 7$.} 
\label{fig:prandtl}
\end{figure}

The variation of the fractal dimension with Prandtl number is shown in Fig.~\ref{fig:dimension}(b). 
The corresponding images of the flow field patterns are shown in Fig.~\ref{fig:prandtl}. 
As the Prandtl number increases the magnitude of the mean flow decreases and eventually the 
spiral defect chaos state vanishes and is replaced with a stationary pattern~\cite{chiam:2003}.
We find that the fractal dimension decreases rapidly with increasing $\sigma$ as shown 
by the solid line in Fig.~\ref{fig:dimension}(b).  For the range $1\leqslant \sigma \lesssim 2$ the solid 
line is a curve fit given by 
\begin{equation}
D_\lambda= \alpha \sigma^{-\beta} - \gamma
\end{equation}
where $\alpha = 44.95$, $\beta=1.06$ and $\gamma=20.91$. This curve fit was determined using 
numerical results in the range $1\leqslant \sigma \leqslant 1.8$ and it predicts the zero of the fractal 
dimension to occur at $\sigma=2.06$. From our numerical results the fractal dimension vanishes to 
within the accuracy of our calculations for $\sigma \geqslant 2$ and is represented by the horizontal solid 
line.  For $\sigma \geqslant 2$ the fluid patterns slowly evolve to a time-independent stationary pattern 
as shown in Fig.~\ref{fig:prandtl}(b)-(d).  Our results suggest that the fractal dimension is inversely proportional 
to the Prandtl number. It is interesting to point out that this is similar to the variation of the mean flow magnitude 
with the Prandtl number as discussed by Chiam \emph{et al.}~\cite{chiam:2003}. 

\section{Conclusions}

A fundamental understanding of high-dimensional chaotic dynamics in spatially extended 
systems remains a vast and important challenge.  In this paper, we have used large scale 
numerics to provide a quantitative link between powerful ideas of dynamical systems theory 
and a fluid system that can be explored in the laboratory.  We have gone to considerable 
computational effort to perform simulations for the geometries, boundary conditions, and 
system parameters that are of experimental relevance. Our computation of the Lyapunov 
based diagnostics provide results that are currently not possible to obtain analytically or 
experimentally and we have used these to provide new physical insights.  Although the 
Lyapunov based diagnostics we have quantified are not directly accessible to experimental 
measurement, at least not in any straightforward way that we can suggest, the values 
we present are an important benchmark for comparison as further experimental and theoretical 
work is conducted. For example, it may be possible to connect our results with experimental 
measurements using ideas based upon Lagrangian coherent structures~\cite{voth:2002,haller:2001} 
or computational homology~\cite{kurtuldu:2011}. From a theoretical point of view, our work 
suggests that it would be interesting to explore the dynamics of the spectrum of Lyapunov 
\emph{vectors} using the more recently suggested approach of characteristic Lyapunov 
vectors that satisfy Oseledec splitting~\cite{ginelli:2007,pazo:2008}. Overall, we anticipate 
that our results will be useful to those interested in controlling, predicting, and modeling 
high-dimensional chaotic systems.

\bigskip

\noindent Acknowledgments: The computations were conducted using the 
resources of the Advanced Research Computing center at Virginia Tech and 
the research was supported by NSF grant no. CBET-0747727.  We have also 
had many fruitful discussions with Mike Cross, Paul Fischer, Janet Scheel, Keng-Hwee 
Chiam, Magnus Einarsson, and Nicholar O'Connor.

\end{document}